\documentclass[superscriptaddress,showpacs,preprintnumbers,amsmath,amssymb,
nofootinbib,twocolumn,aps,prd,10pt]{revtex4-1}
\pdfoutput=1
\usepackage{amssymb,amsmath}
\usepackage{epsfig}
\usepackage{xcolor}
\usepackage{float}
\usepackage[utf8]{inputenc}
\usepackage{stmaryrd}
\usepackage{mathrsfs}
\usepackage{mathalfa}
\usepackage[normalem]{ulem}

\DeclareMathOperator\erf{erf}

\DeclareMathOperator\erfi{erfi}
\DeclareMathOperator\sgn{sgn}
\DeclareMathOperator\Si{Si}
\DeclareMathOperator\Ci{Ci}
\DeclareMathOperator\Ei{Ei}

\newcommand{\A}{\mathcal{A}}

\newcommand{\G}{\mathcal{G}}

\newcommand{\la}{\langle}
\newcommand{\ra}{\rangle}
\newcommand{\be}{\begin{equation}}
\newcommand{\ee}{\end{equation}}
\newcommand{\ba}{\begin{eqnarray}}
\newcommand{\ea}{\end{eqnarray}}
\newcommand{\beg}{\begin{gather*}}
\newcommand{\eng}{\end{gather*}}
\newcommand{\hh}{,\hspace{0.5cm}}
\newcommand{\hhh}{,\hspace{0.2cm}}

\newcommand{\eq}[1]{(\ref{#1})}
\newcommand{\lap}{\triangle}
\newcommand{\n}[1]{\label{#1}}

\newcommand{\ins}[1]{{\mbox{\tiny #1}}}
\newcommand{\ind}[1]{{\mbox{\scriptsize #1}}}

\newcommand{\ts}[1]{{\boldsymbol{#1}}}

\def\XXint#1#2#3{{\setbox0=\hbox{$#1{#2#3}{\int}$ }
\vcenter{\hbox{$#2#3$ }}\kern-.6\wd0}}

\usepackage[makeroom]{cancel}
\usepackage[caption=false]{subfig}
\usepackage[colorlinks=true,
            citecolor=green,
            linkcolor=red,
            filecolor=cyan,
            urlcolor=magenta,
            backref=false]{hyperref}

\newcommand{\dd}{\mbox{d}}

\newcommand{\ff}{$\langle \varphi^2(x)\rangle $}
\newcommand{\rf}{\langle \varphi^2(x)\rangle_{\mbox{\tiny{ren}}} }
\newcommand{\rr}{$\langle \varphi^2(x)\rangle_\ind{ren}$}

\begin{document}

\title{Probing the vacuum fluctuations in scalar ghost-free theories}

\author{Jens Boos}
\email{boos@ualberta.ca}
\affiliation{Theoretical Physics Institute, University of Alberta, Edmonton, Alberta, Canada T6G 2E1}
\author{Valeri P. Frolov}
\email{vfrolov@ualberta.ca}
\affiliation{Theoretical Physics Institute, University of Alberta, Edmonton, Alberta, Canada T6G 2E1}
\author{Andrei Zelnikov}
\email{zelnikov@ualberta.ca}
\affiliation{Theoretical Physics Institute, University of Alberta, Edmonton, Alberta, Canada T6G 2E1}

\date{\today}

\begin{abstract}
We discuss the response of vacuum fluctuations to a static potential in the context of massive, ghost-free infinite-derivative scalar field theories in two dimensions. For the special case of a $\delta$-like potential, $V=\lambda \delta(x)$, the problem is exactly solvable and we calculate the corresponding Hadamard function for this quantum field. Using this exact result we determine the renormalized value of the vacuum polarization $\la \hat{\varphi}^2(x)\ra_\ind{ren}$ as a function of the distance $x$ from the position of the potential. This expression depends on the amplitude of the potential as well as the scale of non-locality $\ell$; for distances $x\gg\ell$ the non-local and local results agree, whereas for distances $x < \ell$ there is a difference.
\end{abstract}

\pacs{03.65.Pm, 03.70.+k, 11.10.Lm \hfill Alberta-Thy-01-19}

\maketitle


\section{Introduction}
The existence of zero-point fluctuations distinguishes a quantum field from a classical field. For a free field in empty flat spacetime these fluctuations are not observable and one usually neglects them. In other words, one considers renormalized quantities in which the contribution of free vacuum zero-point fluctuations is omitted by their subtraction. However, in the presence of matter interacting with the quantum field zero-point fluctuations might lead to observable effects. A famous example is the Casimir effect: The presence of conducting metals and dielectrics changes the propagation of zero-point modes. Their contribution to the vacuum expectation value of the energy is modified, and  this energy depends on the shape and position of macroscopic bodies. Thus, as a result of vacuum fluctuations, there appear forces acting on these bodies. This effect was described by Casimir in 1948 \cite{Casmir:1947hx,Casimir:1948dh}. In 1997 Lamoreaux \cite{Lamoreaux:1996wh} directly measured the force between two closely spaced conducting surfaces to within 5\% and experimentally confirmed the existence of the Casimir effect.

There are different ways to calculate the Casimir force that give the same result \cite{Milton:2004ya,Bordag:2009}. Let us consider two parallel conducting plates. As a result of the fluctuations, there exist microscopic currents in the plates. The average of the (retarded) forces between such currents does not vanish and depends on the distance between the plates, and thereby gives rise to the Casimir force. In the other way of calculation, one can focus on the electromagnetic zero-point fluctuations in the cavity between the plates. Taking the presence of the plates into account by properly choosing boundary conditions for the quantum field then yields the Casimir force. In the second approach one can also calculate the renormalized quantum average of stress-energy tensor $\la \hat{T}_{\mu\nu}\ra$.

Quantum vacuum averages of quantities that are quadratic in the field depend on the boundary conditions and on an external potential or a current. Quite often, these quantum averages are called \emph{vacuum polarization}. Using this terminology, one may say that the Casimir effect is a result of the vacuum polarization produced by conducting plates. Certainly, one can characterize the vacuum polarization by considering  other quantities instead of the stress-energy tensor. For example, for a scalar field $\hat{\varphi}$ one may study the properties of $\la \hat{\varphi}^2\ra$, and one may consider this object as a ``poor man's version of $\la \hat{T}_{\mu\nu}\ra$.'' In the present paper we use this option. Namely, we consider the field $\hat{\varphi}$ obeying the following equation
\be \n{LEQ}
[\hat{{\cal D}}-V(x)]\hat{\varphi}=0\, .
\ee
In order to specify the operator $\hat{{\cal D}}$, let us consider an analytic function ${\cal D}(z)$ of the complex variable $z$. The operator $\hat{{\cal D}}$ is then obtained by substitution $z\to \Box-m^2$.

We consider and compare two different cases. In both cases $V(x)$ is an external potential producing the vacuum polarization. In the first case we put ${\cal D}(z)=z$, such that the operator $\hat{{\cal D}}$
is just  Klein--Gordon operator $\Box-m^2$ and Eq.~\eqref{LEQ} describes a local massive scalar field.

In the second case let us instead consider the function ${\cal D}(z)$ = $z \exp[f(z)]$, where $f(z)$ is an entire function (and therefore has no poles in the complex plane). Then, the inverse of this function
\be
{1\over {\cal D}(z)}\equiv {\exp[-f(z)]\over z}
\ee
has only one pole at $z=0$. This implies that the propagator $1/\hat{{\cal D}}$ does not have ghosts at tree level and hence the theory \eqref{LEQ} has the same number of propagating degrees of freedom as in the first case, which is why these theories theories are called \emph{ghost-free}. Since an exponential of a derivative operator contains infinitely many derivatives by means of its series expansion, these ghost-free theories are also called ``infinite-derivative theories'' or ``non-local theories.'' We use these terms interchangeably.

Later on, we shall consider a special class of ghost-free theories specified by a positive integer number $N$,
\begin{align}
\label{eq:gfn}
f(z)=(-z \ell^2)^N \, ,
\end{align}
which we call $\mathrm{GF}_N$. The parameter $\ell$ is a critical length (or time) at which the modifications connected with the non-locality become important. Technically, this length scale appears in order to form the dimensionless combination $\ell^2(\Box-m^2)$. Let us introduce the symbol
\be
\label{eq:gfn-form}
\alpha(z)=\exp[-f(z)]\, ,
\ee
which we call a \emph{form factor}. These form factors need to have the proper behavior such that we can reproduce the local theory in a certain limit. For this purpose let us consider again the $\mathrm{GF_N}$ class of theories. In a Fourier basis one has $\ell^2(\Box-m^2) \rightarrow \ell^2(\omega^2 - q^2 - m^2)$, where $\omega$ and $q$ denote the temporal and spatial Fourier frequencies, respectively. The local limit is obtained when $\omega\ell \ll 1$, $q\ell \ll 1$, and $m\ell \ll 1$. Hence, in a more general case, it corresponds to the behavior of the differential operator $\hat{\mathcal{D}}(z)$ at $z=0$. Therefore, in order to obtain the correct infrared behavior that reproduces the standard local theory in the limit $z\to 0$, one needs to demand that all physical form factors satisfy $\alpha(0)=1$. This is evidently the case for the class of $\mathrm{GF_N}$ theories \eqref{eq:gfn}, but there are other choices as well.

Ghost-free field theories, and especially ghost-free gravity, have been discussed in a large number of publications, starting from the papers \cite{Tomboulis:1997gg,Biswas:2011ar,Modesto:2011kw,Biswas:2013cha}; see also \cite{Shapiro:2015uxa,Buoninfante:2018gce} for recent developments. The main driving force of the study of such theories is an attempt to improve the ultraviolet behavior of the theory without introducing unphysical (ghost) degrees of freedom. For applications of ghost-free gravity for resolving cosmological as well as black hole singularities, see e.g.\ \cite{Biswas:2010zk,Biswas:2013kla,Conroy:2015nva,Modesto:2017sdr,Buoninfante:2018xiw,Koshelev:2018hpt}; in the context of gravitational waves see \cite{Kilicarslan:2018unm,Kilicarslan:2019njc}.

The main goal of the present paper is to study the properties of zero-point fluctuations in the ghost-free theory. To probe such fluctuations we consider their response to a specially chosen potential $V(x)$. We restrict ourselves to the simplest case when this potential is static and is of the form of a $\delta$-like barrier. We demonstrate that for such a potential both problems, local and non-local one, are exactly solvable. In the main part of the paper we assume that the flat spacetime is two-dimensional. At the end we discuss the higher dimensional versions of the theory, and we also make remarks on the thermal fluctuations in the ghost-free theory in the presence of the potential $V(x)$.

\section{Scalar ghost-free theory}

We begin by considering a simple two-dimensional model of a ghost-free massive scalar field interacting with a potential $V$. We denote Cartesian coordinates by $X=(t,x)$, such that the Minkowski metric is
\be
\dd s^2 = -\dd t^2 + \dd x^2 \, .
\ee
The action of the theory reads
\be\begin{split}\label{S2D}
S&=\frac12 \int \dd^2{X}\ \Big[\varphi\, \hat{{\cal D}} \varphi-V\varphi^2\Big] \, .
\end{split}
\ee
For a quantum field $\hat{\varphi}$ this action gives Eq.~\eqref{LEQ}. The operator $\hat{{\cal D}}$ is a function of the Klein--Gordon operator $\Box-m^2$. Its explicit form for the local and non-local ghost-free theories was discussed in the Introduction. In order to study the vacuum polarization we choose a static potential $V(x)$ that has the form of a simple $\delta$-function
\be\label{V}
V = \lambda\,\delta(x) \, ,
\ee
where we assume that this potential is repulsive such that $\lambda>0$. For the calculations we shall employ the formalism of Green functions. Since there exists a wide set of different Green functions related to our problem, let us first discuss them and introduce notations that will be used throughout the rest of this paper.

\subsection{Green functions ``zoo''}

In general, we denote a Green function as $G(X,X')$ with a different choice of the fonts. For the Green functions in the local theory, in the presence of the potential, we use the bold font
\be
\ts{G}^\bullet(X,X') \, ,
\ee
where $\bullet=\big(+,-,(1),\mathrm{F,R,A}\big)$ denotes the type of the Green function:
\be
\ts{G}^\bullet=\begin{cases}
\ts{G}^{+}& \mbox{positive  frequency Wightman function}\\
\ts{G}^{-}& \mbox{negative frequency Wightman function}\\
\ts{G}^\ind{(1)}& \mbox{Hadamard function}\\
\ts{G}^\ind{F}& \mbox{Feynman propagator}\\
\ts{G}^\ind{R}& \mbox{retarded Green function}\\
\ts{G}^\ind{A}& \mbox{advanced Green function}
\end{cases}
\ee
The first three objects satisfy the homogeneous equation
\begin{align}
[\hat{\mathcal{D}} - V(x)] \ \ts{G}^{+,-,}{}^\ind{(1)}(X,X')=0\, ,
\end{align}
while the last three objects are solutions of the inhomogeneous equation
\begin{align}
[\hat{\mathcal{D}} - V(x)]\ \ts{G}^\ind{F,R,A}(X,X')=-\delta(X-X')\, .
\end{align}
where $\hat{\mathcal{D}} = \Box-m^2$ in this local case.

Similarly, in the non-local ghost-free theory the corresponding Green functions (in the presence of the potential) are denoted by the bold font version of the calligraphic letters
\begin{align}
{\ts{\G}}^\bullet(X,X')\, .
\end{align}
These Green functions obey the equations
\begin{align}
[\hat{\mathcal{D}} - V(x)]\, \ts{\G}^{+,-,}{}^\ind{(1)}(X,X') &= 0 \, , \\
[\hat{\mathcal{D}} - V(x)]\, \ts{\G}^\ind{F,R,A}(X,X') &= -\delta(X-X') \, .
\end{align}
In the absence of the potential, that is, when $V(x)=0$, we shall use for the Green functions the same notations, but without boldface. The expressions
\be
G^\bullet(X-X') \, , \quad \G^\bullet(X-X')
\ee
denote free Green functions in the local and ghost-free theories, respectively.

It should be noted that not every method of quantization of local theories is applicable to the case of non-local theories. There are different approaches toward adapting traditional methods of quantization to non-local ghost-free theories:

For example, the definition of a quantization procedure using Wick rotation from the Euclidean signature to the Lorentz signature may not work for ghost-free theories. However, one may postulate that quantum field theory is well defined only in the Euclidean setup \cite{Buoninfante:2018mre,Asorey:2018wot} and then try to extract information about observables in a physical domain. This approach is attractive from a mathematical point of view because in the Euclidean geometry and in the local case the propagator is unique and well-defined. In the non-local case, however, the propagator picks up essential singularities in several asymptotic directions in the complex momentum plane, rendering the evaluation of correlators and contour integrals impossible.

An alternative approach consists of defining the non-local quantum theory in the physical domain with the Lorentz signature without ever resorting to Wick rotation. In the present paper we accept the second approach, which is technically more involved but conceptually clearer: The quantization employed in the present paper makes use of Green functions as well as their asymptotic boundary conditions which are well known in local field theory. As we will show, in this particular setting (a static, $\delta$-shaped potential) these methods are sufficient to construct a unique non-local quantum theory.

Of course non-locality requires us to reassess some concepts such as local causality and time-ordering. In particular, time-ordering is no longer applicable in the non-local theory in the traditional way. Also, the notions of retarded and advanced propagators are to be properly generalized because local causality in ghost-free theories is not respected. Usually, theories without local causality are prone to instabilities and hence undesirable from a physical point of view. But this generally accepted belief is not necessarily applicable to the entire class of non-local theories: there are some ghost-free theories that are free from any instabilities.

In local theories the retarded (advanced) propagator $G^\ind{R(A)}(x,y)$ vanishes provided $x$ lies everywhere outside the future (past) null cone of the point $y$. As it has been formulated by DeWitt \cite{DeWitt:1965jb}, in non-local theories this boundary condition is to  be replaced by an asymptotic condition that causal propagators vanish only in the ``remote past'' (``remote future''). Similarly, the boundary conditions for the Feynman propagator have to be replaced by the asymptotic conditions. This approach is well-defined and adequate for the computation of various scattering amplitudes in the presence of external potentials, in spite of the fact that there appear some acausal effects in the vicinity of the potential. We shall comment on these conceptual issues in more detail elsewhere \cite{BFZ:upcoming}.

The presence of the potential $V(x)$ breaks the Poincar\'e invariance of the free theory in two ways: first, it violates translational invariance, and second, it selects a reference frame in which the potential is at rest. However, since the potential is static, the model preserves the translation invariance in time. This means that all Green functions depend only on the time difference $t-t'$ of their arguments. This makes it possible and convenient to use the temporal Fourier transformation. For a function $\varphi(t,x)$ we denote
\ba
\varphi_{\omega}(x)&=&\int\limits_{-\infty}^{\infty} \dd t \,e^{i\omega t} \varphi(t,x)\, ,\\
\varphi(t,x)&=&\int\limits_{-\infty}^{\infty} {\dd\omega\over 2\pi} \,e^{-i\omega t} \varphi_{\omega}(x) \, .
\ea
The Fourier transform of the operator $\hat{{\cal D}}$ is
\be\n{FFF}
\hat{{\cal D}}_{\omega}={\cal D}(\partial_x^2+\varpi^2)\hh
\varpi=\sqrt{\omega^2-m^2}\, .
\ee
The temporal Fourier transforms of the above Green functions are marked by the subscript $\omega$:
\be
\ts{G}^\bullet_\omega(x,x')\hhh \ts{\G}^\bullet_\omega(x,x')\hhh G^\bullet_\omega(x-x')\hhh \G^\bullet_\omega(x-x')\, .
\ee
In the presence of the $\delta$-potential the model also has the discrete reflection symmetry $x\to - x$. This implies that
\ba
\ts{G}^\bullet_\omega(x,x')&=&\ts{G}^\bullet_\omega(-x,-x')\, ,\\
\ts{\G}^\bullet_\omega(x,x')&=&\ts{\G}^\bullet_\omega(-x,-x')\, .
\ea

\subsection{Free local and ghost-free Green functions}

Non-local equations are well known in condensed matter theory. For example, the propagation of perturbations in a homogeneous dispersive medium can be described by \eqref{LEQ} with $\hat{{\cal D}}=-\partial_t^2 -f(\lap)$, where $\lap$ is the Laplace operator. Quasiparticles associated with such a theory have the dispersion relation $\omega^2=f(-k^2)$, where $\omega$ is the energy, and $k$ is a momentum of the quasi-particle. A property which distinguishes the ghost-free theory from other non-local theories is that its action is locally Lorentz invariant.
The corresponding dispersion relation is ${\cal D}(-\omega^2+k^2)=0$. This means that any solution of the homogeneous equation \eqref{LEQ} in the local theory is automatically a solution of the homogeneous ghost-free equation. In other words, the on-shell solutions in the local and ghost-free cases are the same.

Let us present now useful expressions for the temporal Fourier transforms of some Green functions which will be used later. We use the following notations:
\ba
\varpi&=&\sqrt{\omega^2-m^2}\ \mbox{  for  } |\omega|\ge m\, ,\\
\kappa&=&\sqrt{m^2-\omega^2}\ \mbox{  for  } |\omega|< m\, .
\ea
For this definition both quantities are real non-negative quantities. Let us also notice that in the absence of the potential $V$ the Green functions (for both the local and non-local cases) depend only on the difference $x-x'$ of their arguments. In what follows we denote this difference simply by $x$.

In the local theory the Hadamard function reads
\be
\label{eq:hdm-free}
{G}^\ind{(1)}_{\omega}(x)=\theta(|\omega|- m)\,{\cos (\varpi\,x)\over \varpi}\, ,
\ee
while the Feynman propagator and  the retarded Green function are
\begin{align}
\label{eq:feynman-free}
{G}^\ind{F}_{\omega}(x)&=\begin{cases} \frac{i}{2 \varpi}\,{e^{i\varpi\,|x|}}, & \ \mbox{  for  } |\omega|\ge m\,;\\
{1\over 2 \kappa}\,{e^{-\kappa\,|x|}},& \ \mbox{  for  } |\omega|< m \, .
\end{cases} \\
{G}^\ind{R}_{\omega}(x)&=\begin{cases}{i\varepsilon_{\omega}\over 2 \varpi}\,{e^{i\varepsilon_{\omega}\varpi\,|x|}},& \ \mbox{  for  } |\omega|\ge m \, ;\\
{1\over 2\kappa}\,{e^{-\kappa\,|x|}},& \ \mbox{  for  } |\omega|< m \, .
\end{cases}
\end{align}
Here and in what follows we denote $\varepsilon_{\omega}=\sgn(\omega)$. As mentioned previously, all these functions are invariant under the change $x\to -x$. If $\omega\ge -m$ one has that
${G}^\ind{R}_{\omega}(x)$ coincides with ${G}^\ind{F}_{\omega}(x)$. For $\omega\ge 0$ the following relation is valid:
\be \n{FD0}
{G}^\ind{(1)}_{\omega}(x) = 2\Im [ {G}^\ind{R}_{\omega}(x) ]\, .
\ee
The last equality is nothing but the fluctuation-dissipation theorem for the vacuum (zero temperature) case, and we shall comment on this in the Conclusion.

Let us now discuss the free Green functions for a generic non-local ghost-free theory.\footnote{A comprehensive discussion of the Green functions in the ghost-free theory can also be found in \cite{Buoninfante:2018mre}.} Note that the discussion which follows is valid for any non-local theory that can be formulated in terms of one form factor $\alpha$. To begin with, in the absence of the potential one has
\be\label{GG}
\G_{\omega}^\ind{(1)}(x)=G_{\omega}^\ind{(1)}(x)\, .
\ee
In local quantum field theory the free Hadamard function is defined as the symmetric expectation value
\begin{align}
\label{eq:def-hdm-qft-free}
\G^{(1)}(X,X') \equiv \la \hat{\varphi}(X)\hat{\varphi}(X')+\hat{\varphi}(X')\hat{\varphi}(X)\ra \, ,
\end{align}
where the expectation value is performed in the vacuum state and readily reproduces Eq.~\eqref{eq:hdm-free}. As seen in Eq.~\eqref{GG}, in the non-local free theory one obtains the same Hadamard function as in the local case. The Feynman propagators and the retarded Green functions in the non-local theory differ from their local versions by a universal term ${\Delta \G}_{\omega}(x)$ as follows:
\begin{align}
\label{eq:deltaG}
\G^\ind{F,R}_{\omega}(x) = G^\ind{F,R}_{\omega}(x)+{\Delta \G}_{\omega}(x)\, .
\end{align}
This additional term is given by the integral
\begin{align}
\Delta\G_{\omega}(x)&=\int\limits_{-\infty}^{\infty}{\dd q\over 2\pi}\cos(q x) {1-\alpha(\varpi^2-q^2)\over \varpi^2-q^2} \, .\n{INT}
\end{align}
Since the form factor $\alpha$ has the property $\alpha(0)= 1$, the integrand is a regular function at $q^2=\varpi^2$. Let us also notice that $\Delta\G_{\omega}(x)$ is a real function which is invariant under the transformation $x\to -x$. Last, in the local case when $\alpha=1$ one has $\Delta\G_\omega(x)=0$.

In what follows, we will recast all our results in terms of this modification term $\Delta\G_\omega(x)$ since it captures the impact of the non-local modification on the local theory.

\section{Green functions in the presence of the potential}
In this part we will derive exact expressions for the Hadamard function as well as the causal propagators (retarded and Feynman) for the ghost-free theory in the presence of the $\delta$-potential.

\subsection{Lippmann--Schwinger equation and its solution}

For the calculation of the response of zero-point fluctuations to an external potential one needs to find the corresponding Hadamard Green function. For our choice of the potential it is possible to obtain it in an explicit form. Consider the equation
\begin{align}
\label{eq:eom-phi-modes}
\hat{ {\cal D}}_{\omega}\varphi_{\omega}(x) -V(x)\varphi_{\omega}(x)=0\, .
\end{align}
Denote by $\varphi^0_{\omega}(x)$ a solution of the equation for $V=0$. Then one can write a solution of \eqref{eq:eom-phi-modes} for the mode function $\varphi_{\omega}(x)$ as
\be
\label{eq:ls-fourier}
\varphi_{\omega}(x)=\varphi^0_{\omega}(x)-\int\limits_{-\infty}^\infty \! \dd x' \, \G^\ind{R}_{\omega}(x,x')V(x') \varphi_{\omega}(x')\, .
\ee
This is a so-called Lippmann--Schwinger equation \cite{Lippmann:1950zz}.

For $V(x)=\lambda \delta(x)$ the integral can be taken explicitly and one obtains
\be
\varphi_{\omega}(x)=\varphi^0_{\omega}(x)-\lambda \G^\ind{R}_{\omega}(x)\varphi_{\omega}(0)\, .
\ee
Here we used that the free Green function $\G^\ind{R}_{\omega}(x,x')$ depends only on the difference of the coordinates $x-x'$; we denote such a function of one variable for $x'=0$ as $\G^\ind{R}_{\omega}(x)$.
Provided $1+\lambda\G^\ind{R}_\omega(0)\not=0$ this algebraic equation can be easily solved and one obtains
\begin{align}
\begin{split}
 \n{FLS}
\varphi_{\omega}(x) & =\varphi^0_{\omega}(x)-\Lambda_\omega \varphi^0_{\omega}(0) \G^\ind{R}_{\omega}(x) \, , \\
\Lambda_\omega & ={\lambda \over 1+\lambda \G^\ind{R}_{\omega}(0)}\, .
\end{split}
\end{align}
Formally one can employ the free advanced Green function $\G^\ind{A}_\omega(x)$ as well, and it will also solve Eq.~\eqref{eq:eom-phi-modes}. Expanding a physical wave packet with ``advanced modes'' instead of ``retarded modes'' will correspond to different boundary conditions. However, we will prove below that both modes give rise to the same Hadamard function.

\subsection{Hadamard function}
The Hadamard function in the $X$-representation is defined as the symmetric expression
\be
\ts{\G}^{(1)}(X,X') \equiv \la \hat{\varphi}(X)\hat{\varphi}(X')+\hat{\varphi}(X')\hat{\varphi}(X)\ra \, ,
\ee
such that $\ts{\G}{}^{(1)}(X,X') = \ts{\G}{}^{(1)}(X',X)$. Applying a temporal Fourier transform results in the expression
\be
\ts{\G}^{(1)}_{\omega}(x,x')=\la \hat{\varphi}_{\omega}(x)\hat{\varphi}_{-\omega}(x')+\hat{\varphi}_{-\omega}(x') \hat{\varphi}_{\omega}(x)\ra\, ,
\ee
and the symmetry of $X \leftrightarrow X'$ implies that
\begin{align}
\label{eq:hdm-fourier-smtry}
\ts{\G}^{(1)}_{-\omega}(x,x') = \ts{\G}^{(1)}_{\omega}(x',x) \, .
\end{align}
These are formal expressions, but due to Eq.~\eqref{GG} we can relate them to local expressions in a unique way: Using Eq.~\eqref{FLS} for the field operator $\hat{\varphi}_{\omega}(x)$ and the property \eqref{GG} one obtains
\begin{align}
\begin{split}
\label{eq:hdm-result}
&\ts{\G}^{(1)}_{\omega}(x,x')\equiv{G}^{(1)}_{\omega}(x-x') \\
&-\Lambda_\omega {\G}^{R}_{\omega}(x) {G}^{(1)}_{-\omega}(x') -\Lambda_{-\omega}{\G}^{R}_{-\omega}(x') {G}^{(1)}_{\omega}(x)\\
&+{G}^{(1)}_{\omega}(0) \Lambda_\omega {\G}^{R}_{\omega}(x) \Lambda_{-\omega}{\G}^{R}_{-\omega}(x')\, .
\end{split}
\end{align}
We take this as a unique prescription for obtaining the non-local, interacting Hadamard function. In the case of vanishing potential, $\lambda=0$, or in the case of vanishing non-locality, $\Delta\G=0$, we recover the local results.

Let us now discuss the properties of relation \eqref{eq:hdm-result}. By construction, this expression satisfies \eqref{eq:hdm-fourier-smtry}. Second, by means of Eq.~\eqref{eq:hdm-free}, it is proportional to $\theta(|\omega|-m)$ and
\begin{align}
\label{eq:hdm-integral-bounds}
\ts{\G}^{(1)}_{\omega}(x,x') = 0 \quad \text{for} \quad |\omega| < m \, .
\end{align}
Last, let us notice that
\begin{align}
\label{eq:hdm-properties-0}
\ts{\G}^{(1)}_{-\omega}(x,x') = \ts{\G}^{(1)}_{\omega}(x,x') \, .
\end{align}
This, combined with \eqref{eq:hdm-fourier-smtry}, finally implies
\begin{align}
\label{eq:hdm-properties}
\ts{\G}^{(1)}_{\omega}(x,x') = \ts{\G}^{(1)}_{|\omega|}(x,x') =  \ts{\G}^{(1)}_{|\omega|}(x',x) \, .
\end{align}
Again, one might substitute the free advanced Green function $\G^\ind{A}_\omega(x)$ in the above relations. It is related to the free retarded Green function via
\begin{align}
\label{eq:relation-free-adv-free-ret}
\Lambda^\ind{A}_\omega = \Lambda_{-\omega} \, , \quad \G^\ind{A}_\omega(x) = \G^\ind{R}_{-\omega}(x) \, ,
\end{align}
where we defined the analogous quantity
\begin{align}
\Lambda^\ind{A} := \frac{\lambda}{1+\lambda\G^\ind{A}_\omega(0)} \, .
\end{align}
Then one may define
\begin{align}
\begin{split}
\label{eq:hdm-result-adv}
&\ts{\G}^{(1)}_{\omega}(x,x')_\ind{A} ={G}^{(1)}_{\omega}(x-x') \\
&-\Lambda^\ind{A}_\omega {\G}^{A}_{\omega}(x) {G}^{(1)}_{-\omega}(x') -\Lambda^\ind{A}_{-\omega}{\G}^{A}_{-\omega}(x') {G}^{(1)}_{\omega}(x)\\
&+{G}^{(1)}_{\omega}(0) \Lambda^\ind{A}_\omega {\G}^{A}_{\omega}(x) \Lambda^\ind{A}_{-\omega}{\G}^{A}_{-\omega}(x')\, ,
\end{split}
\end{align}
but using the relations \eqref{eq:relation-free-adv-free-ret} as well as \eqref{eq:hdm-properties-0} one sees that
\begin{align}
\ts{\G}^{(1)}_{\omega}(x,x')_\ind{A} = \ts{\G}^{(1)}_{\omega}(x,x') \, .
\end{align}
Hence, for the calculation of the vacuum polarization in the static case considered here, the retarded and advanced free Green functions can be used interchangeably.

\subsection{Causal propagators}
In this part, let us denote the causal propagators (Feynman and retarded) by the superscript ``C.'' Let us write the causal propagator in the form
\begin{align}
\ts{\G}^\ind{C}_\omega(x,x')&={\G}^\ind{C}_\omega(x-x')+\A_{\omega}(x,x') \, , \label{GW}
\end{align}
where $\mathcal{A}_\omega(x,x')$ satisfies the equation
\begin{align}
\label{WW}
\big[\hat{\mathcal{D}} - V(x)\big]\A_\omega(x,x')=V(x)\,{\G}^\ind{C}_\omega(x-x') \, .
\end{align}
The solution is given by
\begin{align}
\A_\omega(x,x')&=-\int\limits_{-\infty}^\infty \dd x''\,\ts{\G}^\ind{C}_\omega(x,x'')\,V(x'')\,{\G}^\ind{C}_\omega(x''-x') \, . \label{W1}
\end{align}
One may think of this relation as the version of the Lippmann--Schwinger equation for the causal propagators.
Again, for $V(x) = \lambda\delta(x)$ the above integral can be taken and one finds
\be
\A_\omega(x,x')=-\lambda\, \ts{\G}^\ind{C}_\omega(x,0)\,{\G}^\ind{C}_\omega(x') \, .
\ee

Combining this relation with \eq{GW} one gets
\be
\ts{\G}^\ind{C}_\omega(x,x')={\G}^\ind{C}_\omega(x-x')-\lambda\, \ts{\G}^\ind{C}_\omega(x,0)\,{\G}^\ind{C}_\omega(x') \, .
\ee
For $x'=0$ it reduces to the consistency relation
\be
\ts{\G}^\ind{C}_\omega(x,0)={\G}^\ind{C}_\omega(x)-\lambda \,\ts{\G}^\ind{C}_\omega(x,0)\,{\G}^\ind{C}_\omega(0) \, .
\ee
Provided that $1 + \lambda\G^\ind{C}_\omega(0) \not= 0$, we obtain from this algebraic equation the condition
\be
\ts{\G}^\ind{C}_\omega(x,0)={{\G}^\ind{C}_\omega(x)\over 1+\lambda \,{\G}^\ind{C}_\omega(0)} \, .
\ee
Therefore one finally obtains for the causal propagators
\begin{align}
\label{eq:feynman-result}
\ts{\G}{}^\ind{C}_\omega(x,x') = \G{}^\ind{C}_\omega(x-x') - \lambda\,{{\G}^\ind{C}_\omega(x){\G}^\ind{C}_\omega(x')\over 1+\lambda \,{\G}^\ind{C}_\omega(0)} \, ,
\end{align}
where C=F or C=R for the Feynman or the retarded propagator, respectively. By construction, see Eq.~\eqref{eq:feynman-free}, the Feynman propagator satisfies
\begin{align}
\ts{\G}{}^\ind{F}_{\omega}(x,x') &= \ts{\G}{}^\ind{F}_\omega(x',x) \, , \\
\ts{\G}{}^\ind{F}_{-\omega}(x,x') &= \ts{\G}{}^\ind{F}_\omega(x,x') = \ts{\G}{}^\ind{F}_{|\omega|}(x,x') \, ,
\end{align}
as well as
\begin{align}
\label{eq:feynman-integral-bounds}
\Im \, [ \ts{\G}{}^\ind{F}_\omega(x,x') ] = 0 \quad \text{for} \quad |\omega| < m \, .
\end{align}
The retarded propagator, however, satisfies
\begin{align}
\ts{\G}{}^\ind{R}_{\omega}(x,x') &= \ts{\G}{}^\ind{R}_\omega(x',x) \, , \\
\ts{\G}{}^\ind{R}_{-\omega}(x,x') &= \overline{\ts{\G}{}^\ind{R}_\omega}(x',x) \, ,
\end{align}
where the bar denotes complex conjugation.

\subsection{Interrelation between Hadamard function and causal propagators in the static case}
Having the exact expressions for the Hadamard function \eqref{eq:hdm-result} as well as the causal propagators \eqref{eq:feynman-result} at our disposal, it is straightforward to show that they are related via
\begin{align}
\label{eq:relation-green-functions}
\ts{\G}{}^\ind{F}_\omega(x,x') = \frac12 \left( \ts{\G}{}^\ind{R}_\omega(x,x') + \ts{\G}{}^\ind{A}_\omega(x,x') + i \, \ts{\G}{}^\ind{(1)}_\omega(x,x') \right) \, .
\end{align}
Here, $\ts{\G}{}^\ind{A}_\omega(x,x')$ denotes the advanced propagator which can be defined as
\begin{align}
\ts{\G}{}^\ind{A}_\omega(x,x') \equiv \ts{\G}{}^\ind{R}_{-\omega}(x,x') \, .
\end{align}
This implies that also in the $X$-representation one has
\begin{align}
\begin{split}
\ts{\G}{}^\ind{F}(X,X') &= \frac12 \Big( \ts{\G}{}^\ind{R}(X,X') + \ts{\G}{}^\ind{A}(X,X') \\
                        & \hspace{30pt} + i \, \ts{\G}{}^\ind{(1)}(X,X') \Big) \, .
\end{split}
\end{align}
In particular, one can also show that the Hadamard function $\ts{\G}^\ind{(1)}(X,X')$ and the Feynman propagator $\ts{\G}^\ind{F}(X,X')$ are related via
\begin{align}
\label{eq:relation-1}
 \ts{\G}^\ind{(1)}(X,X')=2\Im\big[ \ts{\G}^\ind{F}(X,X')\big] \, ,
\end{align}
which again is due to the Fourier space relation
\begin{align}
\label{eq:relation-2}
 \ts{\G}^\ind{(1)}_\omega(x,x')=2\Im\big[ \ts{\G}^\ind{F}_\omega(x,x')\big] \, .
\end{align}
Evidently, similar relations hold for $V=0$ as well as in the local theories. We prove these relations in Appendix \ref{app:comparison-hadamard-feynman}. It is important to stress that these interrelations are valid for any non-local modification $\Delta \G_\omega(x)$.

Ultimately, we are interested in calculating the vacuum polarization which is defined in terms of the Hadamard function. The above relations show that it is also possible to perform the computations using the Feynman propagator, and  take the imaginary part only at the end. We will make this more precise in the next section.

\section{Vacuum fluctuations}

\subsection{General expression for \rr }

We are interested in the quantity
\begin{align}
\langle\varphi^2(x)\rangle_\ind{ren} &:= \left.\left[ \left\langle \varphi(X) \varphi(X') \right\rangle_{V\not=0} - \left\langle \varphi(X) \varphi(X') \right\rangle_{V=0} \right]\right|_{X=X'} \nonumber \\
&= \frac12 \left. \left(\ts{\G}^\ins{(1)}(X,X') - {\G}^\ins{(1)}(X-X') \right) \right|_{X=X'} \label{eq:phisquared} \\
&= \frac12 \left[ \ts{\G}^\ins{(1)}(X,X') -{G}^\ins{(1)}(X-X') \right] |_{X=X'} \, . \nonumber
\end{align}

Inserting \eqref{eq:hdm-result} into \eqref{eq:phisquared} and using \eqref{eq:hdm-properties} one obtains
\begin{align}
\begin{split}
\label{eq:phisquared-hadamard}
\langle \varphi^2(x) \rangle_\ind{ren} &= -\int\limits_m^\infty \frac{\dd \omega}{2\pi} \Big[ 2\Lambda_\omega {\G}^\ind{R}_{\omega}(x) {G}^\ind{(1)}_{-\omega}(x) \\
&\hspace{45pt} -{G}^\ind{(1)}_{\omega}(0) \left|\Lambda_\omega {\G}^\ind{R}_{\omega}(x)\right|^2 \Big] \, .
\end{split}
\end{align}
Alternatively, inserting \eqref{eq:feynman-result} into \eqref{eq:phisquared} as well as making use of the interrelation \eqref{eq:relation-2} yields
\begin{align}
\label{eq:phisquared-feynman}
\langle \varphi^2(x) \rangle_\ind{ren} = - \Im \left[ \int\limits_m^\infty \frac{\dd\omega}{2\pi} \lambda\,\frac{\left[{\G}^\ind{F}_\omega(x)\right]^2}{1+\lambda \,{\G}^\ind{F}_\omega(0)} \right] \, .
\end{align}
The integration limits follow directly from Eqs.~\eqref{eq:hdm-integral-bounds} and \eqref{eq:feynman-integral-bounds}, respectively. At first glance these two expressions look quite different, but they are, in fact, identical. This can be shown by using the relations detailed in the previous section, as well as in Appendix \ref{app:comparison-hadamard-feynman}. Using expression \eqref{eq:phisquared-hadamard} it is easy to see that in the absence of the potential barrier, that is, when $\lambda=0$, $\rf=0$ as it should be.

Using Eq.~\eqref{eq:deltaG} we can isolate the terms encoding the non-locality and obtain (after changing the integration variable from $\omega$ to $\varpi$) the following expression:
\begin{align}
\begin{split}
\langle \varphi^2(x) \rangle_\ind{ren} &= \int \limits_0^\infty \frac{\dd\varpi}{4\pi} \frac{\Phi_{\omega}(x)}{\sqrt{\varpi^2+m^2}} \, ,
 \\
\Phi_{\omega}(x) &= \frac{B^2-\cos^2(\varpi x)-2\cos(\varpi x)BC}{1+C^2} \, , \\
B &= 2g_\omega(x) - \sin(\varpi|x|) \, , \label{eq:main-integral} \\
C &= 2g_\omega(0) + 2\varpi/\lambda \, ,  \\
g_\omega(x) &= \varpi\Delta\mathcal{G}_\omega(x) \, .
\end{split}
\end{align}
This is a general expression for the renormalized vacuum polarization for any non-local theory specified by $\Delta\G_\omega(x)$ which enters via the dimensionless quantity $g_\omega(x)$.\footnote{The scattering of a scalar field on a $\delta$-like potential in a ghost-free theory was studied in \cite{Boos:2018kir}. By comparing \eqref{eq:main-integral} with the results of this paper one can conclude that the factor $1/(1+C^2)$ which enters the integral \eqref{eq:main-integral} coincides with the transmission probability $R$.}

In what follows, it is our goal to evaluate this expression in the local case, as well as for various non-local cases.

\subsection{Vacuum polarization in the local theory}
\label{subsec:local}

Let us first consider the vacuum fluctuations in the local theory which was studied earlier; see \cite{Bordag:1992cm,Milton:2004ya} and references therein. In terms of calculational techniques our approach is quite similar to the one employed in \cite{Munoz-Castaneda:2013yga}. In what follows we shall use the results of the local theory for the comparison with the results in the ghost-free models. This will allow us to better understand the effects of the non-locality.

In the local case one has $\Delta\G_\omega(x) = 0$, and hence
\begin{align}
B = -\sin(\varpi|x|) \, , \quad C = \frac{2\varpi}{\lambda} \, .
\end{align}
The integral \eqref{eq:main-integral} then takes the form
\begin{align}
\label{eq:vac-pol-local}
\langle \varphi^2(x) \rangle_\ind{ren}^\ind{loc.} &= \lambda \int\limits_0^\infty \frac{\dd \varpi}{4\pi} \frac{2\varpi\sin(2\varpi|x|) - \lambda\cos(2\varpi x)}{\sqrt{\varpi^2+m^2}(4\varpi^2 + \lambda^2)} \, .
\end{align}
Provided $m > 0$ this integral converges, but it is difficult to evaluate this integral analytically.

For $x=0$ we can calculate \eqref{eq:vac-pol-local} analytically and obtain
\begin{align}
\begin{split}
\langle \varphi^2(0) \rangle_\ind{ren}^\ind{loc.} &= - \int\limits_0^\infty \frac{\dd \varpi}{4\pi} \frac{1}{\sqrt{\varpi^2+\mu^2}} \frac{1}{1 + 4\varpi^2} \\
&= \begin{cases} \displaystyle -\frac{\text{arcosh}\left(\frac{1}{2\mu}\right)}{4\pi\sqrt{1-4\mu^2}} & \text{for~} \mu < \tfrac12 \, , \\[15pt]
\displaystyle -\frac{1}{4\pi} & \text{for~} \mu=\tfrac12 \, , \\[15pt]
\displaystyle -\frac{\text{arccos}\left(\frac{1}{2\mu}\right)}{4\pi\sqrt{4\mu^2-1}} & \text{for~} \mu > \tfrac12 \, .
\end{cases}
\end{split}
\end{align}
where $\mu := m/\lambda$. Note that $\langle \varphi^2(0) \rangle_\ind{ren}^\ind{loc.}$ is always negative, and asymptotically one has
\begin{alignat}{3}
\langle \varphi^2(0) \rangle_\ind{ren}^\ind{loc.} &\rightarrow -\infty \quad &&\text{for} \quad \mu \rightarrow 0 \, , \\
\langle \varphi^2(0) \rangle_\ind{ren}^\ind{loc.} &\rightarrow 0 \quad &&\text{for} \quad \mu \rightarrow \infty \, .
\end{alignat}
The divergence for $\mu\rightarrow 0$ corresponds to the well-known IR divergence for a massless scalar field theory in two dimensions.

\begin{figure}[!htb]%
    \centering
    \includegraphics[width=0.47\textwidth]{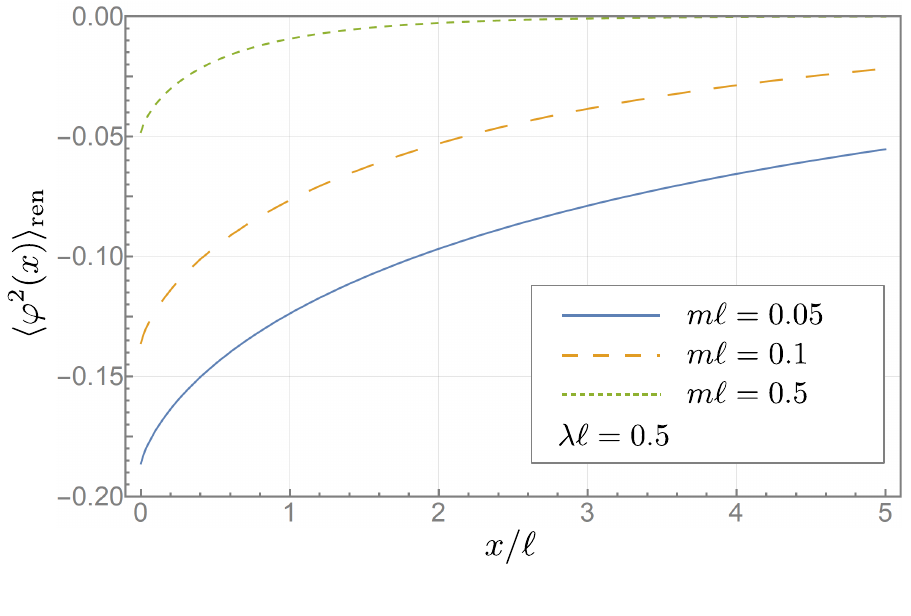}\\[-15pt]
    \caption{The local vacuum polarization \rr\ as a function of the dimensionless distance $x/\ell$ for a fixed potential parameter ($\lambda\ell=0.5$) and for various values of the dimensionless mass parameter $m\ell$.}
    \label{fig:phisquared-loc}
\end{figure}

In the case of $x\not=0$ the vacuum polarization \eqref{eq:vac-pol-local} can be evaluated numerically. In Fig.~\ref{fig:phisquared-loc} we plot the local vacuum polarization $\langle \varphi^2(x) \rangle_\ind{ren}^\ind{loc.}$ as a function of $x$ for different values of the mass $m$.

For the remainder of this paper we shall focus on $\mathrm{GF_N}$ non-local theories for which the non-local modification takes the explicit form
\begin{align}
\label{eq:int-gfn}
\Delta\G_{\omega}(x) &= \int\limits_{-\infty}^\infty\frac{\dd q}{2\pi} \cos(q x) \frac{1-e^{-[\ell^2(q^2-\varpi^2)]^N}}{\varpi^2-q^2} \, .
\end{align}
Note that the integrand is manifestly regular at $q=\varpi$ for all values of $N$. It is also clear that for even $N$ the asymptotic behavior in $\varpi\rightarrow\infty$ is regular, whereas for odd $N$ the asymptotic behavior in $\varpi$ is divergent. This feature will become important in the following discussion.

\subsection{Vacuum polarization in $\mathrm{GF_1}$ theory}

\label{subsec:gf1}
The non-local $\mathrm{GF}_1$ theory is defined by the form factor
\begin{align}
\alpha(z) = \exp(\ell^2 z) \, ,
\end{align}
which is obtained by setting $N=1$ in Eqs.~\eqref{eq:gfn} and \eqref{eq:gfn-form}. In this case the integral \eqref{INT} can be calculated analytically. For $|\omega|\ge m$ (that is, $\varpi>0$) the result is
\begin{align}
\Delta\G_\omega(x) = \frac{1}{2\varpi}\left\{ \sin(\varpi|x|) - \Im \left[ e^{i\varpi |x|} \erf\left( x_+ \right) \right] \right\} \, ,\n{DG}
\end{align}
where we defined
\begin{align}
x_\pm = \frac{|x|}{2\ell} \pm i\omega\ell \, ,
\end{align}
and $\erf(z)$ denotes the error function. In what follows we shall use the fact that the asymptotic of this function for $\Re(z)$=fixed and $\Im(z)\to \pm \infty$ is
\begin{align}
\erf(z) \sim -\frac{e^{-z^2}}{\sqrt{\pi} z} \, .
\end{align}

From expression \eqref{DG} we can read off
\begin{align}
B &= - \Im \left[ e^{i\varpi |x|} \erf\left( x_+ \right) \right] \, , \\
C &= \frac{2\varpi}{\lambda} - \erfi (\varpi\ell) \, ,
\end{align}
where $\erfi(z) = -i\erf(i z)$ denotes the imaginary error function \cite{Olver:2010}. Its asymptotic for real $z\to \infty$ is
\begin{align}
\erfi(z) \sim \frac{e^{z^2}}{\sqrt{\pi} z} \, .
\end{align}
Asymptotically, for finite $\lambda>0$ and $\omega\to \infty$, one has
\begin{align}
B \sim -\frac{1}{\sqrt{\pi}\varpi\ell} e^{\varpi^2\ell^2 - x^2/(4\ell^2)} \hhh
C \sim \frac{1}{\sqrt{\pi}\varpi\ell} e^{\varpi^2\ell^2} \, .
\end{align}
Both of these quantities are exponentially divergent for large frequencies $\varpi$. However, the ratio $B/C$ remains finite in this limit:
\begin{align}
{B\over C}\sim -e^{-x^2/(4\ell^2)}\,,
\end{align}
and one has
\begin{align}
\Phi_\omega(x) \sim e^{-x^2/(2\ell^2)} - 2\cos(\varpi x) e^{-x^2/(4\ell^2)} \, .
\end{align}
The first term in the right-hand side of this expression does not depend on the frequency, and hence the corresponding contribution to \rr\ is logarithmically divergent. By introducing a UV cutoff $\Omega$ one obtains the following expression for the regularized divergent integral:
\begin{align}
Z_0=\int\limits_0^\Omega \frac{\dd\varpi}{4\pi} \frac{1}{\sqrt{\varpi^2+m^2}}={1\over 4\pi}\ln\left({\Omega+\sqrt{\Omega^2+m^2}\over m} \right)\, .
\end{align}
One also has
\begin{align}
Z_1=-\int\limits_0^\infty \frac{\dd\varpi}{4\pi} \frac{2\cos(\varpi x)}{\sqrt{\varpi^2+m^2}}
= -\frac{1}{2\pi} K_0(m|x|) \, ,
\end{align}
where $K_0(x)$ is the modified Bessel function. Using these results one can write the expression for \rr\ in the $\mathrm{GF_1}$ theory as follows:
\begin{align}
\langle \varphi^2(x) \rangle_\ind{ren}^\mathrm{GF_1}&=e^{-x^2/(2\ell^2)} Z_0+e^{-x^2/(4\ell^2)} Z_1+\Psi(x)\, ,\\
 \Psi(x)&=\int \limits_0^\infty \frac{\dd\varpi}{4\pi} \frac{\widetilde{\Phi}_{\omega}(x)}{\sqrt{\varpi^2+m^2}}\, ,\\
 \widetilde{\Phi}_{\omega}(x)&=\Phi_\omega(x) -  e^{-x^2/(2\ell^2)} + 2\cos(\varpi x) e^{-x^2/(4\ell^2)}\, .
\end{align}

\begin{figure}[H]%
    \centering
    \includegraphics[width=0.47\textwidth]{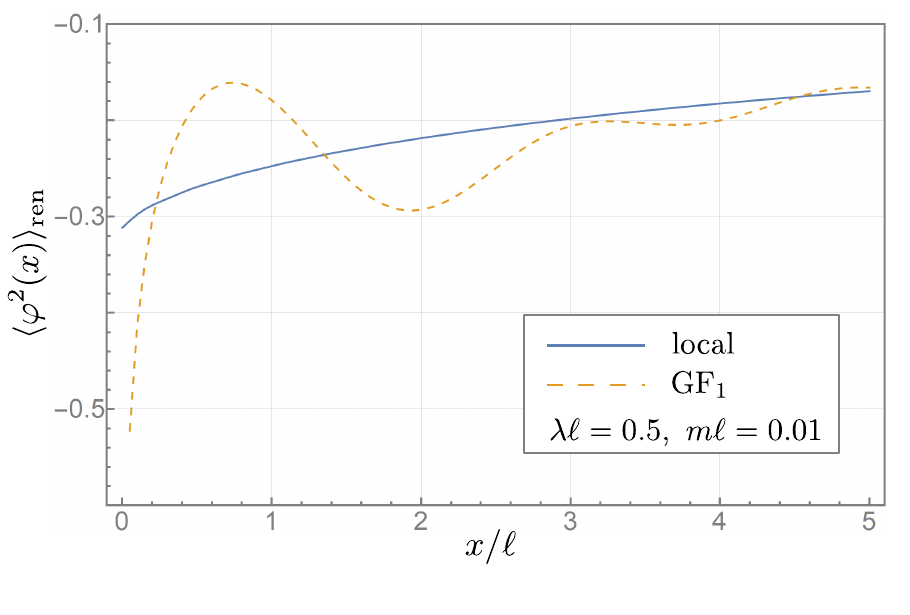}\\[-15pt]
    \caption{We plot the \rr\ in the local case as well as in the $\mathrm{GF_1}$ case (where we subtracted the logarithmically divergent term $Z_0$) as a function of the dimensionless distance $x/\ell$ for a fixed set of potential parameter ($\lambda\ell=0.5$) as well as mass parameter ($m\ell=0.01$). At large distance scales, remarkably, the ``renormalized vacuum polarization'' agrees with the local result. Its shape for small values of $x/\ell$ is drastically different from the local theory.}
    \label{fig:phisquared-loc-gf1}
\end{figure}

The integral for $\Psi(x)$ is convergent. When adding the Bessel function contribution $Z_1$ to $\Psi(x)$ we arrive at some ``renormalized vacuum polarization'' that we can compare to the local expression for \rr. See a graphical comparison of these quantities in Fig.~\ref{fig:phisquared-loc-gf1}.

Our main insights regarding the vacuum polarization in the $\mathrm{GF_1}$ theory are the following: The Gaussian form of the form factor $\alpha(z)$ in this model makes it possible to obtain the Fourier transform of the non-local part of the Green functions \eqref{DG} in an explicit form. This is a very attractive property of this class of ghost-free theories. Namely for this reason, $\mathrm{GF_1}$ theory has been widely used in the study of solutions for static sources. In particular, they effectively regularize the field of a point-like source in four and higher spacetime dimensions (see e.g.\ \cite{Boos:2018bxf} and references therein). However, the propagator in this model behaves poorly in the high-frequency regime, resulting in the peculiar behavior of the field created by a time-dependent source in its near zone (see e.g.\ \cite{Frolov:2016xhq}). In the above calculations of \rr\ we found that the frequency integral for this quantity is logarithmically divergent at high frequencies. The origin of this divergence can easily be traced since the integrand in expression \eqref{eq:int-gfn} exponentially grows when $\varpi\to\infty$. The same property is valid for any $\mathrm{GF_{2n+1}}$ theory, wherein the factor in the numerator grows as $\exp[(\varpi\ell)^{2(2n+1)}]$.

The situation is quite different in the case of $\mathrm{GF_{2n}}$ theories: the corresponding form factor $\alpha(z)$ decreases for both spacelike and timelike momenta when their absolute values tend to infinity. In particular, the integrand in the expression \eqref{eq:int-gfn} exponentially decreases when $\varpi\to\infty$ and is of the order of $\exp[-(\varpi\ell)^{4n}]$. Thus non-local contributions of $\mathrm{GF_{2n}}$ theories are well-defined and divergence-free. However, the analytic calculations in these theories are more involved. In the next section we calculate \rr\ for the $\mathrm{GF_2}$ theory and show that our expectations regarding the finiteness of the vacuum polarization are correct.

\begin{figure}
    \centering
    \includegraphics[width=0.47\textwidth]{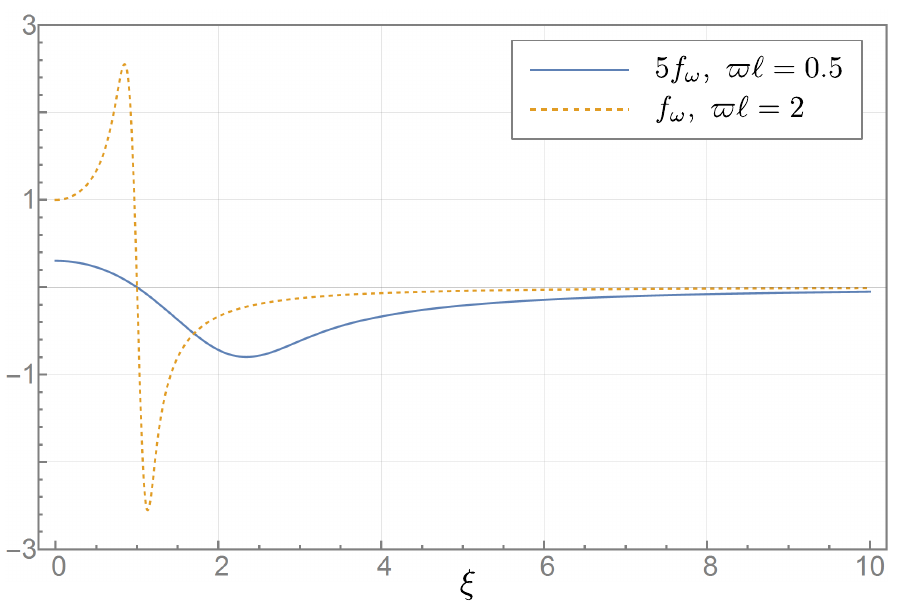}
    \caption{The shape of the function $f_\omega(\xi)$ which enters the integral \eqref{eq:int-gf2} changes drastically for different values of the dimensionless quantity $\varpi\ell$: For small values it is a numerically small smooth function (the solid line in the above plot; to increase visibility we scaled the function by a factor of 5). For larger values of $\varpi\ell$ that surpass the critical value of $\sqrt{2a_2}\approx 1.058\dots$, however, the function begins to vary sharply around $\xi=1$ which fundamentally influences its Fourier transform.}
    \label{fig:plot-fb-1}
\end{figure}

\subsection{Vacuum polarization in $\mathrm{GF_2}$ theory}
\label{subsec:gf2}
The non-local $\mathrm{GF}_2$ theory is defined by the form factor
\begin{align}
\alpha(z) = \exp(-\ell^4 z^2) \, ,
\end{align}
which is obtained from setting $N=2$ in Eqs.~\eqref{eq:gfn}--\eqref{eq:gfn-form}. The non-local modification $g_\omega(x)$ then takes the form
\begin{align}
\label{eq:int-gf2}
g_\omega(x) &= \int\limits_0^\infty\frac{\dd \xi}{\pi} \cos(\xi \tilde{x})f_{\omega}(\xi) \, ,\\
f_{\omega}(\xi)& =\frac{1-e^{-(\varpi\ell)^4(1-\xi^2)^2}}{1-\xi^2} \, ,
\end{align}
where we introduced the dimensionless quantity $\tilde{x}= \varpi x$. We are not aware of any analytic expression for this integral. This property distinguishes this theory from $\mathrm{GF_1}$ theory and necessitates more involved numerical calculations.

It is quite remarkable that for the point at the position of the potential the quantity $g_{\omega}(0)$ can be found analytically. One can use the following representation:
\begin{align}
g_\omega(0) &={1\over 2(\varpi\ell)^2{\pi}^{3/2}}\int\limits_{-\infty}^{\infty}\dd y\, e^{-{y^2\over 4(\varpi\ell)^4}}\int\limits_0^y \dd z P(z) \, ,
\end{align}
where
\begin{align}
P(z)=\int\limits_0^{\infty}\dd\xi\,\sin\left[(1-\xi^2)z\right]={\sqrt{2\,\pi}\over 4\sqrt{z}}\Big(\sin z-\cos z\Big).
\end{align}
The integration over the parameter $z$ and then over $y$ leads to the result
\begin{align}
\begin{split}
g_\omega(0) & = \frac{\sqrt{2} (\varpi\ell)^3}{6\Gamma\left(\tfrac34\right)}~{}_\ins{2}F_\ins{2}\left[\tfrac34,\tfrac54;\tfrac32,\tfrac74;-(\varpi\ell)^4\right]  \\
& -\Gamma\left(\tfrac34\right) \frac{\varpi\ell}{\pi} ~{}_\ins{2}F_\ins{2}\left[\tfrac14,\tfrac34;\tfrac12,\tfrac54;-(\varpi\ell)^4\right] \, .
\end{split}
\end{align}

Let us now consider the case when $x\ne 0$. The integrand in \eqref{eq:int-gf2} contains the function $f_{\omega}(\xi)$; for small values of $\varpi\ell$ it is quite smooth, but for large values of this parameter it has rather sharp features (see Fig.~\ref{fig:plot-fb-1}).
To work numerically, we shall employ a hybrid approach: we approximate the main features of the non-local modification \eqref{eq:int-gf2} analytically and use numerics only for the residual difference between our approximation and the exact expressions (see Appendix \ref{app:gf2} for a detailed explanation of our methods).

We find the following large-$\varpi$ asymptotics:
\begin{align}
g_\omega(0) &= -\frac{1}{4\sqrt{\pi} \varpi^2\ell^2} + \mathcal{O}\left(\varpi^{-6}\right) \, , \\
g_\omega(x) &= \frac{\sin(\varpi |x|)}{2} - \frac{a_2}{3\pi} \left( 2 + e^{-4a_2^2} \right) \frac{x \sin(\varpi x)}{\varpi\ell^2} \nonumber \\
&\hspace{10pt} - \frac{a_2}{2\pi} \left( 3-e^{-4a_2^2} \right)\frac{\cos(\varpi x)}{\varpi^2\ell^2}  + \mathcal{O}\left(\varpi^{-4}\right) \, . \nonumber
\end{align}
Here $a_2$ is a special parameter which we use in our approximation,
\begin{align}
a_2 \approx 0.5604532115\dots \, .
\end{align}
For more details see Appendix \ref{app:gf2}. Thus one obtains the following asymptotic formulas for the parameters $B$ and $C$ which enter \eqref{eq:main-integral} valid in the limit of large values $\varpi$:
\begin{align}
B &\sim -\frac{2a_2}{3\pi} \left( 2 + e^{-4a_2^2} \right) \frac{x \sin(\varpi x)}{\varpi\ell^2} \, , \\
C &\sim \frac{2\varpi}{\lambda} - \frac{1}{4\sqrt{\pi} \varpi^2\ell^2} \, .
\end{align}
The asymptotics for $C$ can readily be reproduced using an alternative analytical approximation scheme, see Appendix \ref{app:gf2n-asymptotics}. As a result we obtain the following asymptotic expression for $\Phi_\omega(x)$ in the limit of large $\varpi$:
\begin{align}
\begin{split}
\Phi_\omega(x) \sim &-\frac{\lambda^2\cos^2(\varpi x)}{4\varpi^2 + \lambda^2}\\
&+ \frac{8a_2\lambda}{3\pi\ell^2}\left(1-e^{-4a_2^2}\right) \frac{x\cos(\varpi x)\sin(\varpi x)}{4\varpi^2 + \lambda^2} \, .
\end{split}
\end{align}

We see that $\Phi_\omega(x)$ is a decreasing function of $\varpi$. Together with the $\sqrt{\varpi^2+m^2}$-factor in \eqref{eq:main-integral} the behavior is improved even more. These considerations imply that---unlike in $\mathrm{GF_1}$ theory---the vacuum polarization for $\mathrm{GF_2}$ theory is well-defined and finite for any value of $x$.

\begin{figure}[!htb]%
    \centering
    \subfloat{{\includegraphics[width=0.47\textwidth]{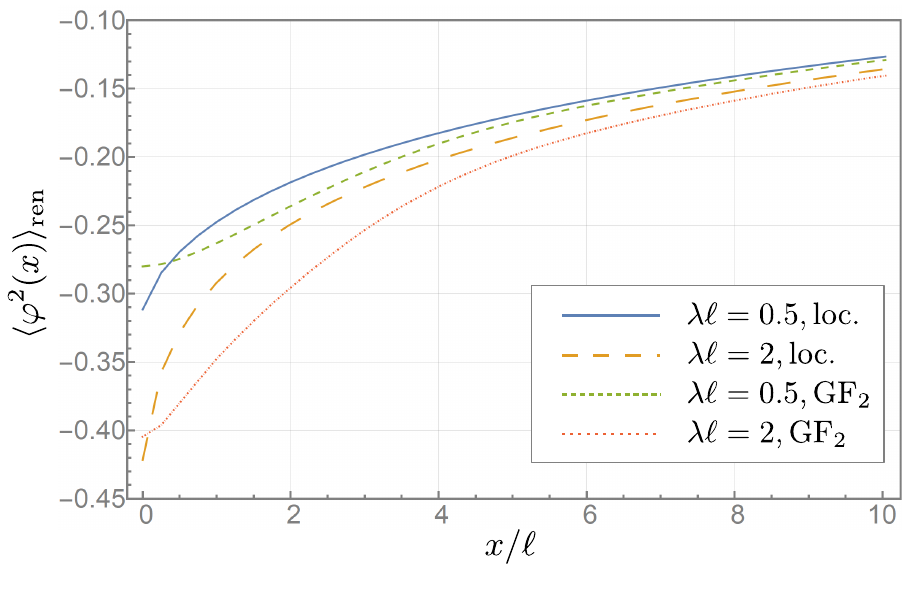} }}\\[-15pt]
    \caption{Local and non-local vacuum polarization $\langle\varphi^2(x)\rangle_\text{ren}$ plotted against the dimensionless distance parameter $x/\ell$ for two different potential parameters ($\lambda\ell=0.5$ and $\lambda\ell=2$). For large distances the local and non-local polarizations approach each other, but for small distance scales $x/\ell \sim \tfrac12$ there is a crossover between the local and non-local vacuum polarization which we previously discussed elsewhere \cite{Boos:2018bhd} on a heuristic level. The effect of the non-locality is a smoothing of the polarization around $x=0$.}
    \label{fig:phisquared-loc-gf2}
\end{figure}

Having a numerical evaluation of $g_\omega(x)$ at our disposal, we can now numerically evaluate $\langle \varphi^2(x) \rangle_\ind{ren}^\mathrm{GF_2}$. The plot of this function (and the comparison to the local theory) can be found in Fig.~\ref{fig:phisquared-loc-gf2}.

\begin{figure}[!htb]%
    \centering
    \subfloat{{\includegraphics[width=0.47\textwidth]{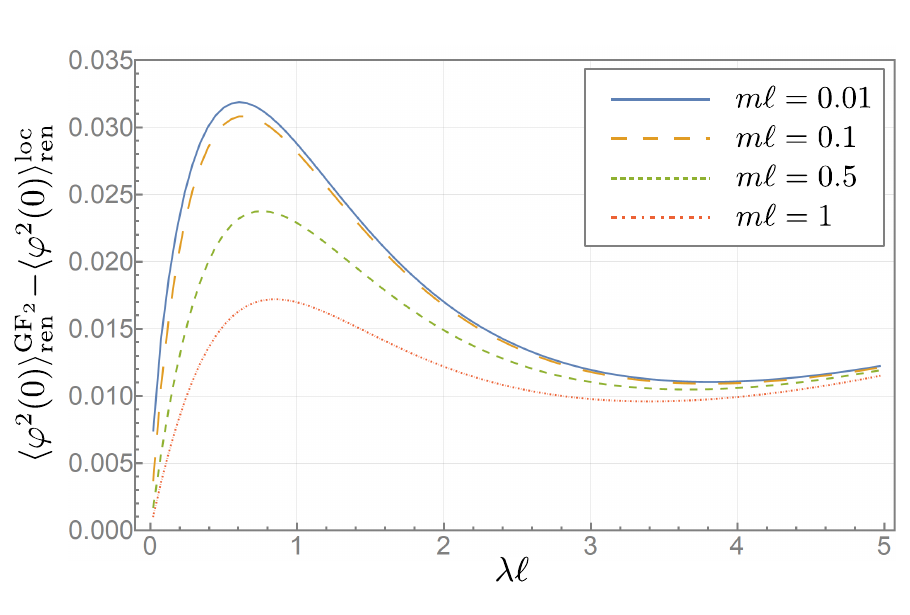} }}\\[-10pt]
    \caption{We plot the difference of the vacuum polarization at the location of the potential at $x=0$ as a function of the potential strength $\lambda\ell$. We see that the difference is a function of the dimensionless mass parameter $m\ell$: for larger masses $m$ at fixed non-locality $\ell$ the difference decreases. In the limiting case $\lambda\rightarrow 0$ the renormalized vacuum polarization vanishes as expected.}
    \label{fig:phisquared0-difference}
\end{figure}

There are a few observations: (i) \emph{Asymptotics.}---For large distances $x\gg\ell$ the vacuum polarization in $\mathrm{GF_2}$ theory approaches that of the local theory, as expected. As this feature is built into all ghost-free theories considered in this paper, this result confirms that our numerical methods work well.
(ii) \emph{Smoothing.}---At small distances scales $x \sim \ell$, however, there is a difference between the local theory and $\mathrm{GF_2}$ theory: the vacuum polarization is smoothed out at the origin $x=0$ as compared to the local case. This implies that all quantities related to the derivative of the vacuum polarization ($\sim \partial_x \varphi^2$) are now regular at the presence of the $\delta$-potential, whereas in the local theory they are not necessarily continuous.
(iii) \emph{Overshoot.}---Across a wide range of masses and potential parameters (quite possible for \emph{all} possible values) the vacuum polarization at the location of the $\delta$-potential is numerically larger than in the local case. We call this an ``overshoot,'' and this feature is plotted in Fig.~\ref{fig:phisquared0-difference}.
(iv) \emph{Crossing.}---Last, at the intermediate location $x\sim\ell$, there is a crossing of the local and $\mathrm{GF_2}$ vacuum polarization. This implies that the difference of the local and non-local vacuum polarization can be both positive and negative. In the $\mathrm{GF_1}$ theory this feature is even more pronounced with multiple crossings, see Fig.~\ref{fig:phisquared-loc-gf1}. We previously discussed these features in the effective energy density in linearized classical non-local gravity \cite{Boos:2018bhd}, and it seems that these crossings or oscillations are a generic feature of ghost-free theories.

In the regularized vacuum polarization obtained in the context of $\mathrm{GF_1}$ theory many of these features appear as well, with the notable exception of point no.~3: the vacuum polarization at the location of the potential is more negative than that of the local theory, which we may call ``undershoot.''

\section{Discussion}
In this paper we discussed a non-local two-dimensional massive scalar quantum field. For the calculations of the vacuum fluctuations of such a field in the presence of a $\delta$-like potential we employed Green-function techniques.

The calculation of \ff \ in the usual local quantum field theory is rather simple. It is greatly simplified by employing a Wick rotation and using the standard methods of the Euclidean theory. In the class of non-local theories which we consider in this paper, however, this method usually does not work: the corresponding form factor $\alpha(z)$ -- see \eqref{eq:gfn-form} -- can infinitely grow when its complex argument $z$ reaches infinity along some directions in the complex plane. As a result, one cannot perform a Wick rotation and all the required calculations are to be done in the ``physical domain'' of the momentum variables. This makes the calculations of the vacuum fluctuations in ghost-free theories much more complicated. In this paper we developed the tools required for these calculations, and this is one of its results.

In order to find \ff \ it is sufficient to obtain the Hadamard Green function. We demonstrated that in the absence of the potential the corresponding Hadamard Green function in the ghost-free theory coincides with a similar function in the local theory. We defined \rr\ as the coincidence limit $x'\to x$ of the difference of the Hadamard Green function of our model and the free local one. This means that \rr\ vanishes in the absence of the potential. However, in the presence of the potential, \rr\ does not vanish in  both non-local and local cases, and the corresponding functions depend on the choice of the theory. The second objective of this paper was to study this effect.

In order to simplify calculations we chose the simple model of a repulsive $\delta$-potential. For such a potential one can find the required Green function in an explicit form by solving the field equations by means of the Lippmann--Schwinger method. The expressions for the Hadamard Green function for a general type of the ghost-free theory as well as integral representations for \rr\  have been obtained in this paper explicitly.

We focused on the calculations of \rr\ for two ghost-free theories ($\mathrm{GF_1}$ and $\mathrm{GF_2}$) and demonstrated that the properties of \rr\ in these models are quite different. In the $\mathrm{GF_1}$ theory the quantity \rr\ is logarithmically divergent, whereas in the $\mathrm{GF_2}$ the quantity \rr\ is a finite smooth function of $x$ for any choice of the mass parameters $m$ and the scale of non-locality $\ell$. The logarithmic divergence of \rr\ in the $\mathrm{GF_1}$ theory is an ultraviolet problem  connected with the behavior of the $\mathrm{GF_1}$ form factor in the high-frequency domain. In the $\mathrm{GF_2}$ theory (as well as for any $\mathrm{GF_{2n}}$ theory) this problem does not exist. For $\mathrm{GF_2}$ theory we also managed to find an exact analytic expression for $\langle\varphi^2(0)\rangle_\ind{ren}$ at the position of the potential. This provided us with a good test of our numerical computations.

We showed that non-local contributions arise from the universal non-local correction term $\Delta\G_\omega(x)$, see Eq.~\eqref{INT}, which is added to the local causal propagators (retarded, advanced, and Feynman). This correction is real-valued and well-defined in the physical Minkowski space for all $\mathrm{GF_N}$ theories.

Our numerical computations demonstrated (see  Figs.~\ref{fig:phisquared-loc-gf1} and \ref{fig:phisquared-loc-gf2}), as we expected, that non-locality smooths the vacuum polarization in the narrow vicinity of the potential and then asymptotically approaches the corresponding value of the local theory. Moreover, at some distance $x < \ell$, there is a crossover between the local and the non-local vacuum polarization. At the location of the potential the ``renormalized'' vacuum polarization of $\mathrm{GF_1}$ is more negative than the local polarization, whereas in the completely regular $\mathrm{GF_2}$ vacuum polarization is larger than the local polarization at $x=0$.

One might think that the model of a two-dimensional massive scalar field, which we consider in this paper, is oversimplified. However, the methods developed here can easily be generalized and adapted to a more realistic case. Suppose that there exist more than one spatial dimension and denote the coordinates in this space as $(x,\vec{y}_{\perp})$. If the potential barrier  still has the form $\lambda\delta(x)$, one can perform the Fourier transform not only with respect to time $t$ but also with respect to transverse coordinates $\vec{y}_{\perp}$. This is possible since the translational invariance into the perpendicular directions is unbroken by the presence of the potential. Denote by $\vec{k}_{\perp}$ the momenta  conjugated to $\vec{y}_{\perp}$. Then one can use the same expression \eqref{FFF} for the operator $\hat{{\cal D}}_{\omega}$ where now  the quantity $\varpi$ takes the form
\begin{align}
\varpi=\sqrt{\omega^2-m^2-\vec{k}^2_{\perp}} \, .
\end{align}
Last, an additional factor depending on $\omega$ appears in the formula \eqref{eq:phisquared-feynman} for \rr, which is connected to the phase volume in momentum space. We hope to address the higher-dimensional problem in a future work.

As a final remark, it would be interesting as well to study the vacuum fluctuations beyond the vacuum state in a thermal bath of finite temperature $T$. An important connected problem lies in studying under which conditions the fluctuation-dissipation theorem is valid in the class of non-local ghost-free theories.

\section*{Acknowledgments}
J.B.\ is grateful for a Vanier Canada Graduate Scholarship administered by the Natural Sciences and Engineering Research Council of Canada as well as for the Golden Bell Jar Graduate Scholarship in Physics by the University of Alberta.
V.F.\ and A.Z.\ thank the Natural Sciences and Engineering Research Council of Canada and the Killam Trust for their financial support.

\appendix

\begin{widetext}
\section{Proof of Equation (\ref*{eq:relation-green-functions})}
\label{app:comparison-hadamard-feynman}
Let us prove the main relation
\begin{align}
\tag{\ref*{eq:relation-green-functions}}
\ts{\G}{}^\ind{F}_\omega(x,x') = \frac12 \left( \ts{\G}{}^\ind{R}_\omega(x,x') + \ts{\G}{}^\ind{A}_\omega(x,x') + i \, \ts{\G}{}^\ind{(1)}_\omega(x,x') \right) \, ,
\end{align}
from which all other relations that me made use of can be derived. Since $\ts{\G}{}^\ind{F}_\omega(x,x') = \ts{\G}{}^\ind{F}_{-\omega}(x,x')$ we can restrict the proof, without loss of generality, to the case of $\omega > 0$. Then one has
\begin{align}
\begin{split}
\label{eq:app:main}
\G{}^\ind{R}_\omega(x) &= \G{}^\ind{F}_\omega(x) = G{}^\ind{F}_\omega(x) + \Delta\G{}_\omega(x) \, , \\
\G{}^\ind{A}_\omega(x) &= \G{}^\ind{R}_{-\omega}(x) = \overline{G{}^\ind{F}_\omega}(x) + \Delta\G{}_\omega(x) \, , \\
\G{}^\ind{(1)}_\omega(x) &= G{}^{(1)}_\omega(x) = -i\left[ G{}^\ind{F}_\omega(x) - \overline{G{}^\ind{F}_\omega}(x) \right] \, .
\end{split}
\end{align}
Note that $\Delta\G{}_\omega(x) \in \mathbb{R}$ as well as $\Delta\G{}_{-\omega}(x) = \Delta\G{}_\omega(x)$. Let us also define (for $\omega>0$)
\begin{align}
\Lambda{}_\omega := \frac{\lambda}{1 + \lambda G^\ind{F}_\omega(0) + \lambda \Delta\G^\ind{F}_\omega(0)} \, ,
\qquad \Lambda_\omega^\ind{R} = \Lambda_\omega \, ,
\qquad \Lambda_{-\omega}^\ind{R} = \overline{\Lambda_\omega} \, .
\end{align}
Now we can express all interacting non-local expressions in terms of the free, local Feynman propagator $G{}^\ind{F}_\omega(x)$ as well as the real-valued modification $\Delta\G_\omega(x)$ and the complex function $\Lambda_\omega$:
\begin{align}
\label{eq:app:aux-1}
\ts{\G}{}^\ind{F}_\omega(x,x') &= G{}^\ind{F}_\omega(x-x') + \Delta \G{}_\omega(x-x') - \Lambda_\omega \left[G^\ind{F}_\omega(x) + \Delta\G_\omega(x)\right]\left[G^\ind{F}_\omega(x') + \Delta\G_\omega(x')\right] \, , \\
\label{eq:app:aux-2}
\ts{\G}{}^\ind{R}_\omega(x,x') &= G{}^\ind{F}_\omega(x-x') + \Delta \G{}_\omega(x-x') - \Lambda_\omega \left[G^\ind{F}_\omega(x) + \Delta\G_\omega(x)\right]\left[G^\ind{F}_\omega(x') + \Delta\G_\omega(x')\right] \, , \\
\label{eq:app:aux-3}
\ts{\G}{}^\ind{A}_\omega(x,x') &= \overline{G{}^\ind{F}_\omega}(x-x') + \Delta \G{}_\omega(x-x') - \overline{\Lambda_\omega}\left[\overline{G^\ind{F}_\omega}(x) + \Delta\G_\omega(x)\right]\left[\overline{G^\ind{F}_\omega}(x') + \Delta\G_\omega(x') \right] \, , \\
i\ts{\G}{}^\ind{(1)}_\omega(x,x') &= G{}^\ind{F}_\omega(x-x') - \overline{G{}^\ind{F}_\omega}(x-x') \nonumber \\
&\hspace{12pt} - \Lambda_\omega \left[G^\ind{F}_\omega(x) + \Delta\G_\omega(x)\right] \left[ G{}^\ind{F}_\omega(x') - \overline{G{}^\ind{F}_\omega}(x') \right] -\overline{\Lambda_\omega} \left[\overline{G^\ind{F}_\omega}(x') + \Delta\G_\omega(x')\right] \left[ G{}^\ind{F}_\omega(x) - \overline{G{}^\ind{F}_\omega}(x) \right]  \nonumber \\
\label{eq:app:aux-4}
&\hspace{12pt} + \Lambda_\omega \left[G^\ind{F}_\omega(x) + \Delta\G_\omega(x)\right] \overline{\Lambda_\omega} \left[\overline{G^\ind{F}_\omega}(x') + \Delta\G_\omega(x')\right] \left[ G{}^\ind{F}_\omega(0) - \overline{G{}^\ind{F}_\omega}(0) \right] \, .
\end{align}
In the last line we can recast the term proportional to $\Lambda_\omega\overline{\Lambda_\omega}$ as follows:
\begin{align}
\Lambda_\omega\overline{\Lambda_\omega}\left[ G{}^\ind{F}_\omega(0) - \overline{G{}^\ind{F}_\omega}(0) \right] = \overline{\Lambda_\omega} - \Lambda_\omega \, .
\end{align}
Then, we can insert the above expressions into \eqref{eq:relation-green-functions}. Comparing the terms independent of $\Lambda_\omega$ as well as linear terms in $\Lambda_\omega$ then yields the identity. Realizing that $\ts{\G}^\ind{R}_\omega(x,x') + \ts{\G}^\ind{A}_\omega(x,x')$ is real-valued, one can take the imaginary part of \eqref{eq:relation-green-functions} and obtain the relation
\begin{align}
\tag{\ref*{eq:relation-1}}
\ts{\G}^\ind{(1)}_\omega(x,x') = 2\Im \left[ \ts{\G}{}^\ind{F}_\omega(x,x') \right] \, .
\end{align}
Using the above expressions \eqref{eq:app:aux-1}--\eqref{eq:app:aux-4} it can now be verified explicitly. The above results holds true for any choice of the form factor; the key assumptions lie in the properties of the local propagators as well as the real-valuedness of the non-local correction $\Delta\G_\omega(x,x')$.

\section{$\Delta\G_\omega(x)$ in $\mathrm{GF_2}$ theory}
\label{app:gf2}
The dimensionless non-local modification $g_\omega(x) = \varpi\Delta\mathcal{G}_\omega(x)$ for $\mathrm{GF_2}$ is given by the integral
\begin{align}
\label{eq:gf2-main-integral}
g_\omega(x) &= \int\limits_0^\infty \frac{\dd \xi}{\pi} \cos(\xi \tilde{x}) f_b(\xi) \, , \quad f_b(\xi) := \frac{1-e^{-b^2(1-\xi^2)^2}}{1-\xi^2} \, , \qquad \tilde{x} := \varpi x \, , \qquad b := (\varpi\ell)^2 \, .
\end{align}
This integral is well-defined, but we are not aware of any analytic solution. For $x=0$, however, there exists a solution. Note that in this section, for numerical convenience, we denote $f_\omega(x)$ as defined in Eq.~\eqref{eq:int-gf2} of the main body of the paper as $f_b(x)$ instead, where $b \equiv (\varpi\ell)^2$.

\subsection{Exact form of $\Delta\G_\omega(0)$}
The function $f_b(\xi)$ as taken from Eq.~\eqref{eq:gf2-main-integral} can be represented as the integral
\be\begin{split}
f_b(\xi)&={1\over 2b\sqrt{\pi}}\int\limits_{-\infty}^{\infty}\dd y\, e^{-{y^2\over 4b^2}} ~{1-\cos[(1-\xi^2)y]\over 1-\xi^2}={1\over 2b\sqrt{\pi}}\int\limits_{-\infty}^{\infty}\dd y\, e^{-{y^2\over 4b^2}}\int\limits_0^y \dd z \sin[(1-\xi^2)z] \, .
\end{split}\ee
Hence
\be\begin{split}
\pi g_{\omega}(\tilde{x}) &= \int\limits_0^\infty \dd \xi\, \cos(\xi \tilde{x}) f_b(\xi)={1\over 2b\sqrt{\pi}}\int\limits_{-\infty}^{\infty}\dd y\, e^{-{y^2\over 4b^2}}\int\limits_0^y \dd z P(z,\tilde{x}) \, ,
\end{split}\ee
where
\be
P(z,\tilde{x})=\int\limits_0^\infty \dd\xi\,\cos(\xi \tilde{x})\sin[(1-\xi^2)z]=-{\sqrt{\pi}\over 2\sqrt{z}}\cos\left({\tilde{x}^2\over 4z}+z+{\pi\over 4}\right) \, .
\ee
For $\tilde{x}=0$ one can take the integrals exactly using the relation
\be
\int\limits_0^y \dd z P(z,0)= -{\pi\over 2}\left[C\left(\sqrt{2y\over\pi}\right)-S\left(\sqrt{2y\over\pi}\right)\right] \, .
\ee
Here $C$ and $S$ are the Fresnel integrals. Then
\be
\pi g_\omega(0)={\sqrt{b}}
\left[{\sqrt{2}\pi b\over 6\Gamma\left(\tfrac34\right)}~{}_\ins{2}F_\ins{2}\left(\tfrac34,\tfrac54;\tfrac32,\tfrac74;-b^2\right)-\Gamma\left(\tfrac34\right)~{}_\ins{2}F_\ins{2}\left(\tfrac14,\tfrac34;\tfrac12,\tfrac54;-b^2\right)\right].
\ee
We find the asymptotics
\begin{align}
\label{eq:gf2-g0-asymptotics}
g_\omega(0) \sim \begin{cases}
\displaystyle -{\Gamma\left(\tfrac34\right)\over{\pi}} \sqrt{b} + \mathcal{O}\left(b^{3/2}\right) &\text{for}~ b \ll 1 \, , \\[15pt]
\displaystyle -{1\over 4\sqrt{\pi}\, b} + \mathcal{O}\left(b^{-3}\right) \quad &\text{for~} b \gg 1 \, .
\end{cases}
\end{align}
See a plot of this function in Fig.~\ref{fig:plot-n}.

\subsection{Semi-analytic approach of calculating $\Delta\G_\omega(x)$}
We are not aware of any analytic solution of \eqref{eq:gf2-main-integral} for $x\not=0$. In what follows, we will describe our method for (i) evaluating this integral numerically and (ii) extracting the asymptotic behavior for large $\varpi$.

Depending on the value of the dimensionless parameter $b$, the function $f_b(\xi)$ takes rather different shapes, see Fig~\ref{fig:plot-fb-1}. Calculating the extrema of $f_b(\xi)$ we find a local maximum at $\xi=0$ and a minimum at $\xi=\xi_+$. Moreover, provided $b$ is large enough, there is another local maximum at $\xi=\xi_-$:
\begin{align}
\xi_\pm = \sqrt{1 \pm \frac{2a_2}{b}} \approx 1 \pm \frac{a_2}{b} \, , \qquad \, a_2 = \frac{1}{2\sqrt{2}} \sqrt{ -1 - 2W_{-1}\left( -\frac{1}{2\sqrt{e}} \right) } = 0.5604532115\dots \, ,
\end{align}
where $W_k(x)$ denotes the Lambert W function. It is clear that the maximum $\xi_-$ only appears if $b>2a_2$. For this reason we shall distinguish the two regimes of $f_b(\xi)$ at the intermediate value $b_0 = 3a_2$. For the regime $b<b_0$ we can improve numerical convergence by subtracting the tail of $f_b(\xi)$ analytically,
\begin{align}
b < b_0 : \quad g &\approx -\frac{e^{-|\tilde{x}|}}{2}  + \int\limits_0^{\xi_\infty} \frac{\dd \xi}{\pi} \cos(\xi\tilde{x}) \left[ \frac{1-e^{-b^2(1-\xi^2)^2}}{1-\xi^2} + \frac{1}{1+\xi^2} \right] + E^<_{b_0, \xi_\infty}(\tilde{x}) \, , \\
E^<_{b_0, \xi_\infty}(\tilde{x}) &= \int\limits_{\xi_\infty}^\infty \frac{\dd \xi}{\pi} \cos(\xi\tilde{x}) \left[ \frac{1-e^{-b^2(1-\xi^2)^2}}{1-\xi^2} + \frac{1}{1+\xi^2} \right] \, ,
\end{align}
where $E^<_{b_0, \xi_\infty}(\tilde{x})$ denotes the error of this approximation. 

On the other hand, for the regime $b>b_0$ it is useful to approximate the peak around $\xi=1$ analytically. The following approximation works well:
\begin{align}
\label{eq:gf2-approx}
f_b(\xi) \approx f^\approx_b(\xi) = \begin{cases}
\displaystyle f_b^1(\xi) = \frac{1-c_1 e^{-b^2(1-\xi)}}{1-\xi^2} & \displaystyle \text{for} ~ \xi \le \xi_- \, , \\[15pt]
\displaystyle f_b^2(\xi) = m\xi + n & \displaystyle \text{for} ~ \xi_- < \xi < \xi_+ \, , \\[15pt]
\displaystyle f_b^3(\xi) = \frac{1-c_3 e^{-b^2(\xi-1)}}{1-\xi^2} & \displaystyle \text{for} ~ \xi > \xi_+ \, .
\end{cases}
\end{align}
This linear interpolation between the maximum and minimum captures the sharp variation of $f_b(\xi)$ around $\xi=1$ quite effectively. The values of the parameters $c_{1,3}$ as well as $m$ and $n$ are chosen such that the jump between linear piece ($\xi_- \le \xi \le \xi_+$) and the left and right side is of order $\mathcal{O}(b^{-2})$:
\begin{align}
c_{1,3} = \exp\left[ -4a_2^2 \mp b^2\left( 1 - \sqrt{1\pm\frac{2a_2}{b}} \right) \right] \, , ~
m = \frac12 \left( 1-e^{-4a_2^2} \right) \left( \frac12 - \frac{b^2}{a_2^2} \right) \, , ~
n = \frac12 \left( 1-e^{-4a_2^2} \right) \left( \frac{b^2}{a_2^2} - 1 \right) \, .
\end{align}
The integral over the approximation \eqref{eq:gf2-approx} can be taken exactly. The linear integral is elementary, and for the others one obtains for the indefinite integrals ($\tilde{x}=0$)
\begin{align}
\begin{split}
\int \frac{\dd \xi}{\pi} f^{1,3}_b(\xi) &= \frac{1}{2\pi}\ln\left(\frac{1+\xi}{1-\xi}\right) + \frac{c_{1,3}}{2\pi} \left\{ \Ei[\mp b^2(1-\xi)] - e^{\mp 2b^2} \Ei[\pm b^2(1+\xi)] \right\} \, .
\end{split}
\end{align}
as well as ($\tilde{x}\not=0$)
\begin{align}
\begin{split}
\int \frac{\dd \xi}{\pi} \cos(\xi \tilde{x}) f^{1,3}_b(\xi) &= \frac{\cos{\tilde{x}}}{2\pi} \left\{ \Ci[\tilde{x}(1+\xi)] - \Ci[\tilde{x}(1-\xi)] \right\} + \frac{\sin{\tilde{x}}}{2\pi}\left\{ \Si[\tilde{x}(1+\xi)] - \Si[\tilde{x}(1-\xi)] \right\} \\
&\hspace{12pt}+\frac{c_{1,3}}{2\pi} \Re \left\{ e^{\pm i\tilde{x}}\Ei[\mp(b^2+i\tilde{x})(1-\xi)] - e^{\mp 2b^2} e^{-i\tilde{x}} \Ei[\pm(b^2+i\tilde{x})(1+\xi)] \right\} \, .
\end{split}
\end{align}
In the above $\Si$, $\Ci$, and $\Ei$ denote the sine integral, cosine integral, and exponential integral, respectively:
\begin{align}
\Si(x) := \int\limits_0^x \! \dd t \, \frac{\sin t}{t} \, , \quad \Ci(x) := \gamma +\ln x + \int\limits_0^x \! \dd t \, \frac{\cos t-1}{t} \, , \quad
\Ei(x) := \gamma + \ln x + \int\limits_{-x}^0 \!\dd t \frac{1-e^{-t}}{t} \, .
\end{align}
For the numerical integration we can now write
\begin{align}
b>b_0: \quad g &\approx \int\limits_0^\infty \frac{\dd \xi}{\pi} \cos(\xi\tilde{x}) f^\approx_b(\xi) + \int\limits_0^{\xi_\infty} \frac{\dd\xi}{\pi} \cos(\xi\tilde{x}) \left[ \frac{1-e^{-b^2(1-\xi^2)^2}}{1-\xi^2} - f^\approx_b(\xi) \right] + E^>_{b_0, \xi_\infty}(\tilde{x}) \, , \\
E^>_{b_0, \xi_\infty}(\tilde{x}) &= \int\limits_{\xi_\infty}^\infty \frac{\dd\xi}{\pi} \cos(\xi\tilde{x}) \left[ \frac{1-e^{-b^2(1-\xi^2)^2}}{1-\xi^2} - f^\approx_b(\xi) \right] \, ,
\end{align}
where again $E^>_{b_0, \xi_\infty}(\tilde{x})$ denotes the error of the approximation. See Fig.~\ref{fig:gf2-numerics-2} for a visualization of this numerical integration scheme. There is a subtlety connected with the analytic integral in the case $b>b_0$ because it involves various properties of the cosine integral as well as the exponential integral. Explicitly one obtains ($\tilde{x}=0$)
\begin{align}
\begin{split}
\int\limits_0^\infty \frac{\dd \xi}{\pi} f_b^\approx(\xi) &= \int\limits_0^{\xi_-} \frac{\dd \xi}{\pi} f_b^1(\xi) + \int\limits_{\xi_-}^{\xi_+} \frac{\dd \xi}{\pi} f_b^2(\xi) + \int\limits_{\xi_+}^\infty \frac{\dd \xi}{\pi}  f_b^3(\xi) \\
&=\hspace{10pt}\frac{1}{\pi}\Big\{ \frac{m}{2}(\xi_+^2-\xi_-^2) + n(\xi_+-\xi_-) \Big\} -\frac{1}{2\pi}\left\{ \ln\left(\frac{\xi_+-1}{\xi_++1}\right) - \ln\left(\frac{1-\xi_-}{1+\xi_-}\right) \right\} \\
&\hspace{11pt}+\frac{c_1}{\pi}\Big\{ \Ei\left[-b^2(1-\xi_-)\right] - e^{-2b^2}\Ei\left[b^2(1+\xi_-)\right] - \Ei\left(-b^2\right) + e^{-2b^2}\Ei\left( b^2\right) \Big\} \\
&\hspace{11pt}-\frac{c_3}{\pi}\Big\{ \Ei\left[b^2(1-\xi_+)\right] - e^{2b^2}\Ei\left[-b^2(1+\xi_+)\right] \Big\} \, ,
\end{split}
\end{align}
as well as ($\tilde{x}\not=0$)
\begin{align}
\begin{split}
\int\limits_0^\infty \frac{\dd \xi}{\pi} \cos(\xi\tilde{x}) f_b^\approx(\xi) &= \int\limits_0^{\xi_-} \frac{\dd \xi}{\pi} \cos(\xi\tilde{x}) f_b^1(\xi) + \int\limits_{\xi_-}^{\xi_+} \frac{\dd \xi}{\pi} \cos(\xi\tilde{x}) f_b^2(\xi) + \int\limits_{\xi_+}^\infty \frac{\dd \xi}{\pi} \cos(\xi\tilde{x}) f_b^3(\xi) \\
&=\hspace{10pt} \frac{\cos\tilde{x}}{2\pi}\Big\{ \Ci\left[\tilde{x}\left(1+\xi_-\right)\right] - \Ci\left[\tilde{x}\left(1-\xi_-\right)\right] - \Ci\left[\tilde{x}\left(1+\xi_+\right)\right] + \Ci\left[\tilde{x}\left(\xi_+-1\right)\right] \Big\} \\
&\hspace{11pt}+\frac{\sin\tilde{x}}{2\pi}\Big\{ \pi + \Si\left[\tilde{x}\left(1+\xi_-\right)\right]-\Si\left[\tilde{x}\left(1-\xi_-\right)\right] - \Si\left[\tilde{x}\left(1+\xi_+\right)\right] + \Si\left[\tilde{x}\left(1-\xi_+\right)\right] \Big\} \\
&\hspace{11pt}+\frac{1}{\pi\tilde{x}^2}\Big\{ m\left[\cos\left(\tilde{x}\xi_+\right)-\cos\left(\tilde{x}\xi_-\right)\right] + (m\xi_+ + n)\tilde{x}\sin\left(\tilde{x}\xi_+\right) - (m\xi_- + n)\tilde{x}\sin\left(\tilde{x}\xi_-\right) \Big\} \\
&\hspace{11pt}+\frac{c_1}{2\pi}\Re\Big\{ e^{i\tilde{x}}\Ei\left[(-b^2-i\tilde{x})(1-\xi_-)\right] - e^{-i\tilde{x}-2b^2}\Ei\left[(b^2+i\tilde{x})(1+\xi_-)\right] \Big\} \\
&\hspace{11pt}-\frac{c_1}{2\pi}\Re\Big\{ e^{i\tilde{x}}\Ei\left(-b^2-i\tilde{x}\right) - e^{-i\tilde{x}-2b^2}\Ei\left(b^2+i\tilde{x}\right) \Big\} \\
&\hspace{11pt}-\frac{c_3}{2\pi}\Re\Big\{ e^{-i\tilde{x}}\widetilde{\Ei}\left[(b^2+i\tilde{x})(1-\xi_+)\right] - e^{-i\tilde{x}+2b^2}\widetilde{\Ei}\left[-(b^2+i\tilde{x})(1+\xi_+)\right] \Big\} \, ,
\end{split}
\end{align}
where we defined
\begin{align}
\widetilde{\Ei}(z) := \begin{cases} \Ei(z) \quad &\text{for~} \Re(z) \ge 0 \, , \\
\Ei(z) + i\pi &\text{for~} \Re(z) < 0 \, , \end{cases}
\end{align}
which implements the branch cut of the exponential integral for arguments with negative real part.

\begin{figure*}[!htb]

    \centering
    \subfloat{{\includegraphics[width=0.47\textwidth]{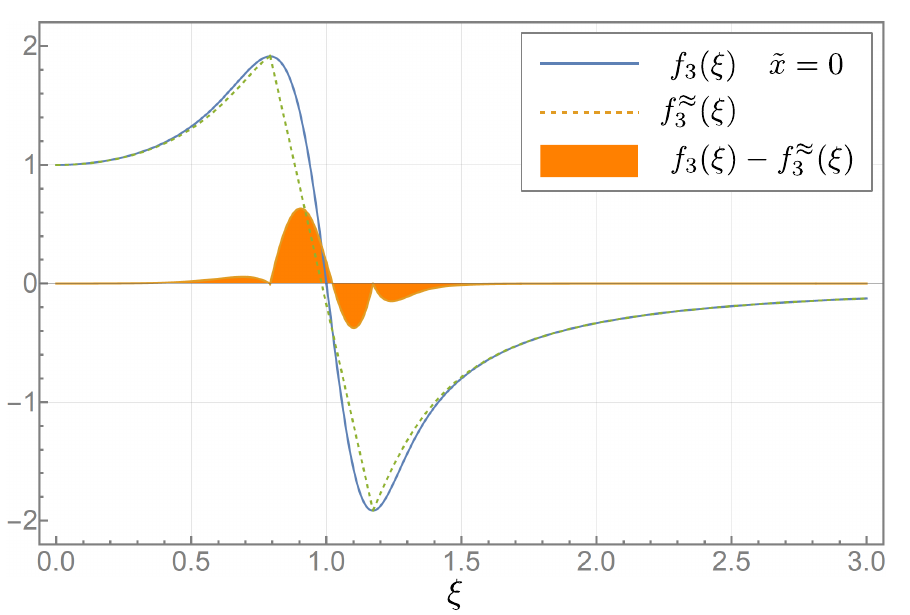} }}%
    \qquad
    \subfloat{{\includegraphics[width=0.47\textwidth]{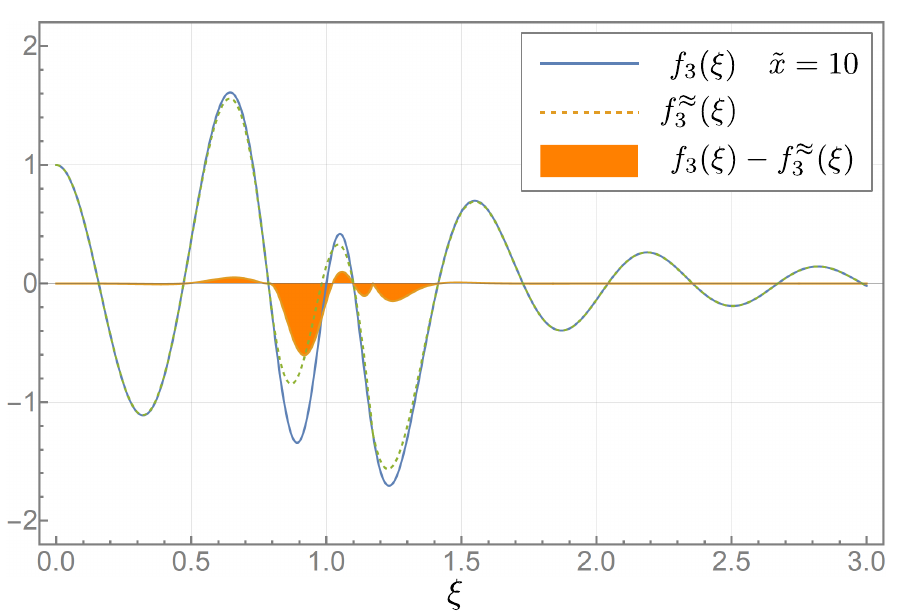} }}
    \caption{Numerical integration scheme: the subtraction of the approximative function $f^\approx_b(\xi)$ (dashed line) from the exact expression (solid line) improves the falloff behavior of the integrand drastically, such that again the numerical integration can be performed in a finite range of $\xi$. In the above, the contributions to the numerical integrals are visualized as the shaded area, and it is clear that the range of non-zero contributions to the numerics is finite for $\tilde{x}=0$ as well as for $\tilde{x}\not=0$.}
    \label{fig:gf2-numerics-2}

    \centering
    \subfloat{{\includegraphics[width=0.47\textwidth]{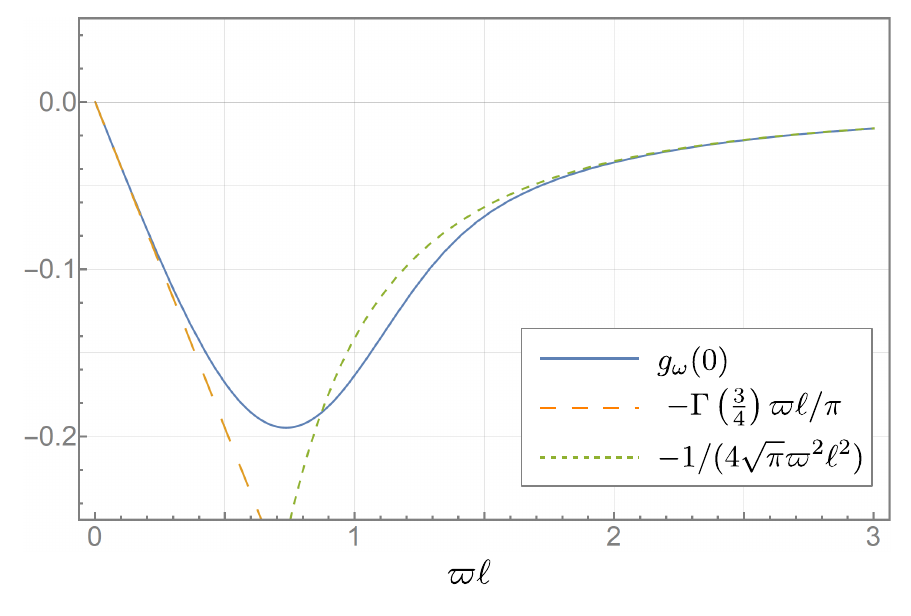} }}%
    \qquad
    \subfloat{{\includegraphics[width=0.47\textwidth]{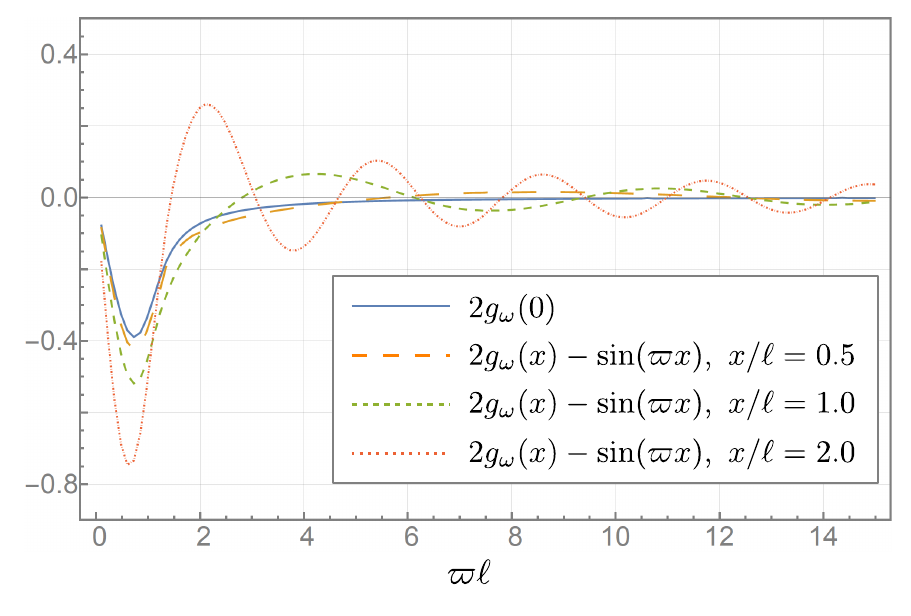} }}
    \caption{The non-local modification $g_\omega(x)$ for $\mathrm{GF_2}$ theory for various values of $x$. Left: Analytic result for $x=0$. Right: Numerical results for various values of $x$, where we subtracted the leading-order oscillating terms $\sin(\varpi x)$. It is visible that the remainder is a decreasing function of $\varpi$.}
    \label{fig:plot-n}

\end{figure*}

\subsection{Asymptotics $x \not= 0$}
Using the approximation presented in Eq.~\eqref{eq:gf2-approx} we can extract the behaviour of \eqref{eq:gf2-main-integral} for large values of $b$ and find (for fixed $\tilde{x}$) the rather crude approximation
\begin{align}
g_\omega(x) &\approx \frac{\sin\tilde{x}}{2} - \frac{a}{2\pi b} \left( 3-e^{-4a_2^2} \right)\cos(\tilde{x}) - \frac{a}{3\pi b} \left( 2 + e^{-4a_2^2} \right) \tilde{x} \sin(\tilde{x}) + \mathcal{O}\left(b^{-2}\right) \, .
\end{align}
The above implies that $g_\omega(x)$ behaves as an oscillatory term of magnitude $\tfrac12$ for large values of $\varpi$. In the above, we made use of the relations
\begin{align}
\begin{split}
\Si(x\rightarrow 0) &\approx x \, , \quad
\Si(x\pm\epsilon) \approx \Si(x) \pm \frac{\sin x}{x}\epsilon \, , \quad
\Ci(x\pm\epsilon) \approx \Ci(x) + \frac{\cos x}{x}\epsilon \, , \\
\Si(x\rightarrow\infty) &\approx \frac{\pi}{2} - \frac{\cos x}{x} \, , \quad
\Ei(x\rightarrow\pm\infty) \approx \pm\frac{e^{\pm x}}{x} \, .
\end{split}
\end{align}
See Fig.~\ref{fig:plot-n} for a graphic confirmation of these asymptotics.

\section{Remarks on \rr\ in $\mathrm{GF_{2n}}$ for larger $n$}
\label{app:gf2n-asymptotics}
As it turns out it is possible to analyze the large$-\varpi$ asymptotics of $g_\omega(0)$ in the case of $\mathrm{GF_{2n}}$ theories for any $n$, which we will describe in the following. Consider again the expression \eqref{eq:gf2-main-integral} in the case of $\mathrm{GF_{2n}}$ theory for $x=0$:
\begin{align}
g_\omega(0) &= \int\limits_{-\infty}^\infty \frac{\dd \xi}{2\pi} f_\beta(\xi) = \frac{1}{2\pi} g(\beta) \, , \quad f_\beta(\xi) := \frac{1-e^{-\beta(1-\xi^2)^{2n}}}{1-\xi^2} \, , \qquad \beta := (\varpi\ell)^{4n} \, .
\end{align}
Now we can calculate
\begin{align}
\partial_\beta g(\beta) = \int\limits_{-\infty}^\infty \dd\xi (1-\xi^2)^{2n-1} e^{-\beta(1-\xi^2)^{2n}} \approx h_- + h_+ \, ,
\end{align}
where the approximation stems from expanding the exponential around its maxima at $\xi_\pm = \pm 1$ for large values of $\beta$. We introduce new variables $y_\pm := \xi \mp 1$ and find
\begin{align}
h_\pm &= \int\limits_{-\infty}^\infty \dd y_\pm (-y_\pm^2 \mp 2y_\pm)^{2n-1} \exp\left[ -\beta(-y_\pm^2 \mp 2y_\pm)^{2n} \right] \\
&= \int\limits_{-\infty}^\infty \dd z_\pm (1+z_\pm) (2z_\pm)^{2n-1} e^{-\beta(2z_\pm)^{2n}} \\
&\approx \int\limits_{-\infty}^\infty \dd z_\pm z_\pm (2z_\pm)^{2n-1} e^{-\beta(2z_\pm)^{2n}} = \frac{2^{2n-1}}{\alpha^{\frac{2n+1}{2n}}}\frac{\Gamma\left(\frac{2n+1}{2n}\right)}{n} \, , \quad \alpha := 2^{2n}\beta = [2(\varpi\ell)^2]^{2n} \, ,
\end{align}
where $z_\pm := \mp y_\pm - \frac12 y_\pm^2$. This implies, by integrating over $\beta$, that one has
\begin{align}
g_\omega(0) \approx \frac{1}{4\pi n} \Gamma\left(\frac{2n+1}{2n}\right)\left( c - \frac{2n}{\varpi^2\ell^2} \right)
\end{align}
In the case $n=1$ we arrive at $\mathrm{GF_2}$ theory and we can compare the above to Eq.~\eqref{eq:gf2-g0-asymptotics}. Inserting $n=1$ into the above one has in the limit $\varpi\ell \gg 1$
\begin{align}
g_\omega(0) \approx \frac{1}{4\pi}\Gamma\left(\frac32\right)\left(c - \frac{2}{\varpi^2\ell^2}\right) = \frac{1}{8\sqrt{\pi}}\left(c - \frac{2}{\varpi^2\ell^2}\right) \, .
\end{align}
In the above considerations, $c$ is an undetermined constant of integration. Setting $c=0$, however, the above exactly reproduces the analytic $\mathrm{GF_2}$ asymptotics of Eq.~\eqref{eq:gf2-g0-asymptotics}, which we consider a somewhat non-trivial consistency check of our approximation schemes.

\end{widetext}

\bibliography{Ghost_references}{}

\end{document}